\catcode`\@=12
\documentstyle[12pt]{article}
 \newcommand{\be}{\begin{eqnarray}}
\newcommand{\BE}{\begin{eqnarray}} \newcommand{\en}{\end{eqnarray}}
\newcommand{\EN}{\end{eqnarray}} \newcommand{\bea}{\begin{eqnarray*}}
\newcommand{\eea}{\end{eqnarray*}} \newcommand{\non}{\nonumber}
\newcommand{\no}{\noindent}
\newcommand{\vs}{\vspace}

\newtheorem{Th}{Theorem}
 \newtheorem{Lem}{Lemma}
\newtheorem{Pro}{Proposition}
\title{{\bf High Temperature Expansions and Dynamical Systems}}

\author{J.Bricmont\thanks{Supported by EC grants SC1-CT91-0695 and
CHRX-CT93-0411}\\ UCL,
Physique Th\'eorique, B-1348, Louvain-la-Neuve,
Belgium\\bricmont@fyma.ucl.ac.be\and
A.Kupiainen\thanks{Supported by NSF grant DMS-9205296 and EC grant
CHRX-CT93-0411} \\
Helsinki University, Department of Mathematics,\\ Helsinki 00014,
Finland\\ajkupiai@cc.helsinki.fi}

\date{}

\begin{document}

\maketitle \begin{abstract}

We develop a resummed high-temperature expansion for lattice spin systems
with long range
interactions, in models where the free energy is not, in general, analytic.
We establish
uniqueness of the Gibbs state and exponential decay of the correlation
functions. Then, we
apply this expansion to the Perron-Frobenius operator of  weakly coupled
map lattices.
\end{abstract}

\section{Introduction.}

The theory of Gibbs states was originally developed for the mathematical
analysis of
equilibrium statistical mechanics. An interesting application of the theory
was found by
Sinai, Ruelle and Bowen in the 70's \cite{Si,Si2,Ru,Bo} who applied it to
the ergodic
theory of uniformly hyperbolic dynamical systems. While this so called
thermodynamic
formalism has been very successful in ergodic theory, the Gibbs states that
describe the
statistics of such dynamical systems are quite simple from the point of view of
statistical mechanics: they describe one dimensional spin systems with
spins taking values
in a finite set and interacting with exponentially decaying potentials. In
particular,
phase transitions, i.e. the coexistence of several Gibbs states for the
same interaction,
which are of major interest in statistical mechanics, are absent in such
systems.

More recently, it has been realized that certain infinite dimensional
dynamical systems
possess attracting sets that are {\it extensive} in a suitable volume. This
is believed to
be the case for many classes of  nonlinear parabolic partial differential
equations on
some spatial domain: the dimension of the attracting set (or a bound for
it) increases to
infinity as the domain becomes unbounded \cite{Te}. Discrete time dynamical
systems, such
as coupled maps, were introduced to model these phenomena \cite{Ka}.
Bunimovich and Sinai
\cite{BS} showed that these systems give rise to a thermodynamic formalism
for spin
systems on a lattice of more than one dimension. Because of this last
feature, the
possibility of phase transitions is at least open. However, most rigorous
analyses have so
far been limited to weakly coupled maps which corresponds to ``high
temperature" in
statistical mechanics. But, even in that regime, coupled maps give rise to
interactions of
a rather peculiar kind, and proving the absence of phase transitions for
the latter is not
completely trivial.

In statistical mechanics, the common wisdom is that ``high temperature" leads
to
uniqueness and decay of correlations (mixing for lattice translations).
Mathematically,
one introduces a suitable Banach space of ``interactions" parametrizing the
Gibbs states
and high temperature means small norm. For small norm the interaction
uniquely determines
the Gibbs state \cite{Do1,Do2}. However, the question of a suitable norm turns
out to be
subtle. One can distinguish between three properties:  uniqueness of the
Gibbs state,
exponential mixing and analyticity (in the potentials) of the correlation
functions. In
the last case one has a complete control of the Gibbs state in terms of a
convergent
expansion, the high temperature expansion. In the other two cases, less
detailed
information is available.

As we shall recall in Section 2, there exist Banach spaces of interactions
where
exponential mixing holds, but analyticity does not hold, in general. It
turns out that the
interactions corresponding, in the thermodynamic formalism, to coupled maps
belong to such
``pathological" Banach spaces of interactions.

The purpose of this paper is twofold. First we wish to give a unified
treatment of the
high temperature states discussed above. We show how all the previous
results (e.g. the
uniqueness results of Dobrushin \cite{Do1,Do2})  and also the
generalizations needed for
dynamical systems can be obtained by a simple resummation of a high
temperature expansion
based on an idea of von Dreifus, Klein and Perez \cite{VKP} used for
disordered systems
(such a resummation was used already by Fisher \cite{Fi}). The resummation
uses in an
essential way the fact that  the interactions are real and thus is not in
conflict with
the lack of analyticity.

Then we apply these results to the study of the ergodic theory of coupled
$C^{1+\delta}$
circle maps. We rederive, generalize and complete the previous studies of
such maps. We
believe this is necessary due to some confusion and incorrect results in
the litterature
(see Section 4). Our approach is slightly novel since we derive the
thermodynamic
formalism without introducing Markov partitions and symbolic dynamics
(which however exist
in the models we consider). One motivation for this is that systems where
one expects
phase transitions (based on numerical and theoretical evidence
\cite{Po,MH}) will, most
probably, not have useful Markov partitions. In our formalism such systems
(e.g. coupled
bounded variation interval maps) nevertheless give rise to a thermodynamic
formalism, but
with potentials whose thermodynamic limit we are presently unable to control.

The paper is organized as follows. Section 2 contains a review of the
theory of Gibbs
states and of the various spaces of potentials. We also formulate and prove
in Section 2 our
main statistical mechanics theorem. This Section is completely independent
of the
dynamical system part of the paper. Section 3 defines coupled map lattices
and states the
main theorem, whose proof can be found in Section 4. We have tried to be rather
self-contained in the statistical mechanics hoping the paper to be
accessible to the
dynamical systems community.

\setcounter{equation}{0} \section{Resummed high-temperature expansions.}

\vs{5mm}

\no {\bf 2.1. Lattice systems}

\vs{5mm}

We consider a lattice spin system: to each $i \in {\bf Z}^d$ we assign
$\Omega_i$, a copy
of a finite set $\Omega_0$ or more generally a compact metric space; we
stick to the
former, but the generalization to the latter case is straightforward. Given
$X \subset
{\bf Z}^d$, a spin configuration in $X$, denoted by $s_X$, is an element
$s_X\in \Omega_X \equiv \times_{i\in X} \Omega_i $.  An interaction is defined
by a family $\Phi = (\Phi_X)$ of (continuous) functions indexed by finite
subsets $X$ of ${\bf Z}^d$: \be \Phi_X : \Omega_X \to {\bf R} \en

We let $\|\Phi_X\|$ denote the sup norm of $\Phi_X$. We consider
translation invariant
$\Phi's: \Phi_{X+i} = \tau_i \Phi_X$ where $\tau_i$, $i \in {\bf Z}^d$ is
the natural
${\bf Z}^d$ action on functions defined on $\Omega_X$ to functions defined on
$\Omega_{X+i}$. Given $\Lambda \subset {\bf Z}^d$, $|\Lambda| < \infty$, and a
configuration $ s'_{\Lambda^c} = (s'_i)_{i\in {\Lambda^c} }$ in
$\Omega_{\Lambda^c}$, the
Hamiltonian in $\Lambda$ (with boundary conditions $ s'_{\Lambda^c}$) is
defined as \be
{\cal H}( s_\Lambda| s'_{\Lambda^c}) = -\sum_{X \cap \Lambda \neq
\emptyset} \Phi_X ( s_{X
\cap \Lambda} \vee s'_ {X \cap \Lambda^c}) \en where, if $ s_X \in \Omega_X$, $
s'_Y \in
\Omega_Y$, for $X\cap Y= \emptyset$, $ s_{X } \vee s'_Y $ is the obvious
configuration in
$\Omega_{X \cup Y}$. The associated (finite volume) Gibbs measure is a
probability
distribution on $\Omega_\Lambda$: \be \nu ( s_\Lambda | s'_{\Lambda^c}) =
Z^{-1} (\Lambda|
s'_{\Lambda^c}) \exp (-{\cal H} ( s_\Lambda| s'_{\Lambda^c})) \en (we put a
minus sign in
(2) for convenience), with \be Z(\Lambda| s'_{\Lambda^c}) = \sum_{
s_\Lambda} \exp (-{\cal
H} ( s_\Lambda| s'_{\Lambda^c})). \en See e.g. \cite{Ru,S,VFS} for more
details on the
theory of Gibbs states.

\vs{5mm}

\no {\bf 2.2. Finite range interactions}

\vs{5mm}

Suppose first that $\Phi_X$ is of finite range, i.e., for some $R < \infty$,
$\Phi_X = 0$ if the diameter of $X$, $d(X) > R$ (we put on ${\bf Z}^d$ the
metric $|i|=\max_\alpha |i_\alpha|$). Then, if $\sup_X \|\Phi_X\|$ is small
enough, a convergent high-temperature expansion (so-called polymer expansion)
yields the following results (see \cite{GM,B,MM,S}): \begin{enumerate}
\item[1)] The Gibbs state is unique i.e., the
finite-volume Gibbs measures (3) converge, as $\Lambda \uparrow {\bf Z}^d$,
independently
of the boundary conditions to a unique measure on $\Omega_{{\bf Z}^d}$
satisfying certain
consistency conditions (the DLR-equations).
\item[2)] The correlation functions in that Gibbs
measure decay exponentially (see (12) below).
 \item[3)] The free energy $F(\Phi)$ defined
as $\lim_{\Lambda \to \infty} |\Lambda|^{-1}\log Z(\Lambda|
s'_{\Lambda^c})$ and the
correlation functions are analytic: Given finite subsets $X_1$,...,$X_n$ and
$\lambda\in{\bf C}^n$, define $\Phi^\lambda$ by $\Phi^\lambda_{Y}=
\lambda_\alpha \Phi_Y$
for each $Y=X_\alpha + j$, for some $j \in {\bf Z}^d$ and some $\alpha =
1,\cdots,n$ and
$\Phi^\lambda_{Y}= \Phi_Y$ for the other $Y$'s. Then, $F(\Phi^\lambda)$ is
analytic in the
polydisc $|\lambda_\alpha|\leq \epsilon$, for $\epsilon$ small enough.
\end{enumerate}

The smallness of $\| \Phi_X \|$ is however not a necessary condition for
uniqueness and
analyticity. For example, one dimensional finite range systems always
satisfy all
three conditions above. Dobrushin-Shlosman \cite{DS1,DS2} gave
several equivalent conditions to be satisfied by the finite volume
distributions (3) that
guarantee that the Gibbs state is unique and that analyticity and
exponential decay of
correlations hold (they call these properties ``complete analyticity").
These results were
later rederived by Olivieri and Picco \cite{O,OP} using expansion methods.

To formulate one of these conditions, used in \cite{O,OP}, it is convenient
to cover $
{\bf Z}^d$ by disjoint cubes of side $L$, called $L$-cubes. We fix such a
covering. We
choose $L$ sufficiently large, and, in particular $L>R$, where $R$ is the
range of
$\Phi^0$.  Then the condition becomes:\\ There exists a function $f:{\bf N}
\to {\bf R}$
such that \be | \frac{Z(\Lambda | s^{xy}) Z(\Lambda | s)}{Z(\Lambda | s^x)
Z(\Lambda |
s^y)} - 1 | \leq f(|x-y|) \en and \be \lim_{n \to \infty} n^{2(d-1)}
f(n)=0. \en Here,
$\Lambda$ is an arbitrary union of $L$-cubes and $s$ an arbitrary
configuration in
$\Omega_{\Lambda^c}$;  $s^A_z = s_z $ for $z \not\in A$, where $A=x,y$ or
$xy$. (Actually,
it is enough to check (5) for $\Lambda$'s being subsets of a sufficiently
large volume.
Then, the fact that (5) holds for larger volumes follows from (13) below.)

\vs{5mm}

\no {\bf 2.3. Infinite range interactions}

\vs{5mm}

Many of these results extend to infinite range interactions but one has to
carefully
distinguish between different Banach spaces of interactions in which
different results are
valid. Without trying to be exhaustive, one has basically the following
results: (see
\cite{S} for a review) \begin{enumerate} \item[1)] If \be \| \Phi \|_1 =
\sum_{0 \in X}
|X| \| \Phi_X \| \en is small enough, then the Gibbs state is unique
\cite{Do1,Do2,Si2}
\item[2)] If, for some $\gamma > 0$, \be \| \Phi \|_2 = \sum_{0\in
X}e^{\gamma d(X)}
\|\Phi_X\| \en is small enough, then the Gibbs state is unique and its
correlation
functions decay exponentially \cite{Gr1}. \item[3)] If, for some $\gamma
>0$, \be \| \Phi
\|_3 = \sum_{0 \in X} e^{\gamma|X|} \| \Phi_X \| \en is small enough, then
the Gibbs state
is unique and its correlation functions are analytic \cite{Is}. \end{enumerate}

The reasons for these different norms are as follows: usually, correlation
functions do
not decay faster than the interactions, so that something like (8) is
needed to get
exponential decay.

As for analyticity, Dobrushin and Martirosyan \cite{DM2} have shown, by
explicit
counterexamples, that analyticity will not hold, (at least not uniformly in
the volume,
see \cite{VFS}, p.958), in a neighbourhood of zero, in any space larger
than the one
defined by (9): Let\\ \be \| \Phi \|_h = \sum_{0\in X} h (|X|) \| \Phi_X\|
. \en Then, if
$\lim_{n \to \infty} h(n) e^{-\gamma n} = 0$, $\forall \gamma > 0$, there
exist (complex)
interactions with arbitrarily small $\| \Phi \|_h$ such that the
corresponding partition
function vanishes for a sequence of cubes $\Lambda_n \to \infty$. (see
\cite{VF}, p.971
for a simple such counterexample).

{}From the point of view of polymer expansions,  the norm (9) is quite natural,
as was remarked by Brydges (\cite{B} p.141): the exponential weight in (9),
depending on the {\it size} of $X$, is needed to control the sum over small
polymers lying inside a big one.

However, this norm is stronger than (7) and, sometimes, than (8): if the
lattice dimension $d$ is larger than one, and if $X$ is, e.g., a large cube,
$|X| >> d(X)$. So, for some $\Phi$'s, uniqueness and exponential decay hold
but analyticity does not, and it seems that the first two properties cannot be
proven using convergent polymer expansions (the existing proofs use
Dobrushin's methods). The problem is that these expansions tend, when they
converge, to yield analyticity almost automatically.

The reader should not be misled by the fact that standard high-temperature
expansions {\it
for spin $\frac{1}{2}$ systems} \cite{GM,Is,CO} seem to show analyticity
for interactions
that are small in a weaker norm than (9). Indeed, in these papers, the
interactions are
written in a special representation (``spin" or ``gas" language) and the
norm used for the
interactions is quite different from the sup norm. This slightly confusing
point is very
nicely clarified in \cite{VF}.

\vs{5mm}

\newpage

\no {\bf 2.4. The main result}

\vs{5mm}

The analyticity results of Dobrushin and Shlosman were extended, for the
analyticity part,
by Dobrushin and Martirosyan \cite{DM1} to infinite range interactions of
the form \be
\Phi = \Phi^0 + \Phi^1 \en where $\Phi^0$ has finite range and is
completely analytic,
while $\| \Phi^1\|_3$ is small (actually, Dobrushin and Martirosyan discuss
the $\Phi^0$
part in terms of specifications instead of interactions; this  could be
done here too). In
the dynamical systems problem of Section 3 we will encounter an interaction
of even more
general type, namely of the form (11), but with $\| \Phi^1\|_2$ (instead of $\|
\Phi^1\|_3$) small. Our main result is

\begin{Th} Let $\Phi = \Phi^0 + \Phi^1$, where $\Phi^0$ is completely
analytic. Then,
there exist $\epsilon >0$, $m>0$, $C< \infty$ such that, if $\| \Phi^1 \|_2
\leq \epsilon$,
there is a unique Gibbs state $\mu$ for $\Phi$ and the correlation
functions satisfy, for
all $F$, $G$, with $F : \Omega_A \to {\bf R}$, $G : \Omega_B \to {\bf R}$,
where $A,B
\subset {\bf Z}^d$ are finite: \be | \langle FG \rangle - \langle F \rangle
\langle G
\rangle | \leq C\|F\|\|G\| e^{-md({A},{B})}, \en where $d({A},{B})$ is the
distance
between the sets $A$ and $B$ and $\langle F \rangle =\int F d\mu$. \end{Th}

\vspace*{5mm}

\par\noindent {\bf Remark 1.} We shall discuss various extensions and
variations of this
result after giving the proof in Section 2.5.

\par\noindent {\bf Remark 2.} Our proof is based on a high-temperature
expansion instead of
Dobrushin's method. We shall resum the expansion and use then, in an
essential way, the
fact that $\Phi_X$ is real, so that our proof applies even when analyticity
does not hold.
It is inspired by a recent work of von Dreifus, Klein and Perez \cite{VKP}
who developed a
high-temperature expansion for disordered systems. Their problem was
similar to ours: due
to Griffiths' singularities \cite{Gri}, analyticity does not hold, in
general, for
disordered systems, while we have to circumvent the possibility of
Dobrushin-Martirosyan
singularities \cite{DM2}. Actually, a simple version of our method appears
already in
Fisher's upper bound on the two-point function for the Ising model
\cite{Fi}. On the other
hand, in a recent paper, Jiang and Mazel \cite{Ma} consider real
interactions similar to
ours, and develop a convergent high-temperature expansion which yields
uniqueness and
exponential decay of the correlation functions, but for systems defined on  a
two-dimensional lattice (their method crucially depends on the fact that in
two dimensions
boundaries of volumes are proportional to their diameters).

\vs{5mm}

\no{\bf 2.5. The proof}

\vs{5mm}

Before starting with the proof, we will use another representation for the
partition
function of $\Phi^0$ derived from (5, 6) by Olivieri and Pico. They show
that if (5, 6)
holds then the interaction $\Phi^0$ ``has a cluster (or polymer)
expansion". Precisely,
this means that the partition function admits the following representation.
If $\Lambda$
is a union of disjoint $L$-cubes, and $ s \in \Omega_{\Lambda^c}$, \be
Z^0(\Lambda | s ) =
e^{f|\Lambda|}\prod_{\alpha} W ( s_{\Lambda^c_\alpha}) \exp \sum_Y \phi_Y( s_{Y
\cap\Lambda^c}) \en where $Z^0$ refers to the partition function with
interaction
$\Phi^0$, the product runs over connected components of $\Lambda^c=\cup_\alpha
\Lambda^c_\alpha$ and the sum runs over connected sets of $L$-cubes $Y$, so
that $Y\cap
\Lambda \neq \emptyset$ (and $\phi_Y$ is a constant if $Y \cap \Lambda^c =
\emptyset$). We
define a subset $X\subset {\bf Z}^d$ to be connected if $\forall i,j \in X$
there exists a
path $i(1),\cdots,i(\ell)$ with $i(1) = i, i (\ell) = j, i(k) \in X, |i(k)
- i(k+1)| = 1,
\forall k,$ where $|i| = \max_{\alpha} |i_\alpha|$. $W \geq 0$ depends on
$s_i$, for $i\in
\Lambda^c_\alpha$, such that $d(i,\Lambda) \leq R$. We have the following
bound: for any
$\epsilon > 0$, there exist $L < \infty$, $\gamma >0$ such that, for the
expansion defined
with $L$-cubes, \be \sum_{0 \in Y} \exp(\frac{\gamma d(Y)}{L}) \| \phi_Y \|
\leq \epsilon
\en (there is no loss of generality in assuming that $\gamma $ here is the
same as in
(8)).

\vspace{5mm} \par\noindent {\bf Remark.} Our condition (5) is Condition A
in \cite{O}
(together with (3.7) in $d=2$), or Condition C in \cite{OP} (see (2.65)). The
representation (13) follows from (2.53), (2.56), (2.60)
in \cite{OP}, or, more precisely,
from the extension of (2.53) to arbitrary boundary conditions. In
\cite{O,OP} volumes of
various shapes are used, but one can always regroup terms and index them by
$L$-cubes, as
was done here for simplicity of notations. The reader may notice that (13)
is what we get
at high temperatures, the only difference being that a site is replaced
here by an
$L$-cube: $f$ is the bulk free energy, the product over the $W$'s gathers
all the boundary
terms, and the sum over $\phi_Y$ is the usual cluster expansion. The
constant $L$ is, in
effect, of the order of the correlation length of the system. Note also
that, if $\Phi^0$
is a finite range one-dimensional interaction (i.e. it couples spins only
along lines
parallel to a lattice axis), then the transfer matrix formalism implies
that (5) holds. In
our application to coupled maps, $\Phi^0$ will be exactly of that form (see
Proposition 3
in Section 4).

\vs{3mm}

\par\noindent {\bf Remark} In the proofs, we shall denote by $C$ or $c$ a
constant that
may vary from place to place.

\vspace*{3mm} \par\noindent {\bf Proof of Theorem 1.} We shall prove (12)
where the
expectation will be taken with respect to a finite volume Gibbs state with
open boundary
conditions (to simplify notations), i.e. the sum in (2) is restricted to $X
\subset
\Lambda$. The corresponding expectation is written $\langle \cdot
\rangle_\Lambda$. Our
bounds are uniform in $\Lambda$. We can then study the limit $\Lambda
\uparrow {\bf Z}^d$,
and prove (12) in that limit. The uniqueness of the Gibbs state can be
shown in a similar
way, and will be discussed at the end of the proof.

First, we write the LHS of (12) using duplicate variables: \be &&2(\langle FG
\rangle_\Lambda - \langle F \rangle_\Lambda \langle G
\rangle_\Lambda)\nonumber \\ &=&
\sum_{ s^{i}_\Lambda,i=1,2} (F( s^{1}_\Lambda) - F( s^{2}_\Lambda))(G(
s^{1}_\Lambda)- G(
s^{2}_\Lambda)) \nu ( s^{1}_\Lambda) \nu ( s^{2}_\Lambda) \nonumber \\ &=&
Z(\Lambda)^{-2}
\sum_{ s^{i}_\Lambda,i=1,2} \tilde F \tilde G \exp (-{\cal H}( s^1_\Lambda)
- {\cal H}(
s^2_\Lambda)) \en where $\tilde F = F ( s^{1}_\Lambda) - F( s^{2}_\Lambda),
\tilde G = G(
s^{1}_\Lambda) - G ( s^{2}_\Lambda)$.

We may replace $\Phi^1_X$ by $\Phi^1_X - \inf_{ s_X} \Phi^1_X$, by adding a
constant to
the Hamiltonian. Thus, we may, without loss of generality, assume that
$\Phi^1_X \geq 0$
for all $X$, and that $ \Phi^1$ still satisfies $\| \Phi^1 \|_2 \leq 2
\epsilon$. Now, we
perform a usual high-temperature expansion on the $\Phi^1$ part of ${\cal
H}$: \be \exp
(\sum_{X,i=1,2} \Phi^1_X ( s^{i}_X)) = \sum_{\cal X} \prod_{X \in {\cal X}}
f_X \en where
the sum runs over sets ${\cal X}$ of subsets of $\Lambda$, and \be f_X =
\exp (\Phi^1_X (
s^1_X) + \Phi^1_X ( s^2_X)) - 1 \en satisfies: \be 0 \leq f_X, \en \be
\sum_{0\in
X}e^{\gamma d(X)} \| f_X \| \leq C \epsilon \en (Use (8) and $\| \Phi^1
\|_2 \leq
\epsilon$). Insert (16) in (15) and, for each term in (16), define $V
=V({\cal X}) =
\underline{A} \cup \underline{B} \cup \underline{{\cal X}}$, where
$\underline{\cal
X}=\cup_{X \in {\cal X} } \underline{X} $ and for any $X\subset {\bf Z}^d$,
$\underline{X}$ is the set of $L$-cubes intersected by $X$.  We have \be (15)=
Z(\Lambda)^{-2} \sum_{\cal X} \sum_{ s^{i}_V,i=1,2} \tilde F \tilde G
\prod_{X \in {\cal
X}} f_X \exp (-{\cal H}^0_V) \prod_{i=1,2} Z^0 (\Lambda \backslash V |
s^{i}_V) \en where
${\cal H}^0_V=-\sum_{X \subset V} (\Phi^0_X ( s^1_X) + \Phi^0_X ( s^2_X))$,
and $ Z^0
(\Lambda \backslash V | s^{i}_V)$ is the partition function with
interaction $\Phi^0$, $
s^{i}_V$ boundary condition in $V$, and open boundary conditions in
$\Lambda^c$. Now, we
use (13) for $Z^0 (\Lambda \backslash V | s^{i}_V)$, and we  define $\Psi_Y
( s_{Y\cap
V})$ by: \be \phi_Y ( s_{Y\cap V}) = \Psi_Y ( s_{Y\cap V}) + \bar \phi_Y
\en where \be
\bar \phi_Y = \min_{ s_{Y\cap V}} \phi_Y ( s_{Y\cap V}) \en so that $\Psi_Y
\geq 0$. Then,
we perform a high-temperature expansion of $\Psi_Y $: \be \exp
(\sum_{Y,i=1,2} \Psi_Y (
s^i_{Y\cap V})) = \sum_{\cal Y} \prod_{Y\in {\cal Y}} g_Y \en where $Y \cap
(\Lambda
\backslash V) \neq \emptyset$ and ${\cal Y}$ is a set of such $Y$'s;
$g_Y=\exp ( \Psi_Y
(s^1_{Y \cap V}) + \Psi_Y (s^2_{Y\cap V}))-1$ satisfies: \be 0 \leq g_Y \en
\be \sum_{0
\in Y} \exp(\frac{\gamma d(Y)}{L})  \| g_Y \| \leq C \epsilon \en (using
(14)). Finally,
we insert the result in (20). We get \be (20)= Z(\Lambda)^{-2} \sum_{{\cal
X},{\cal Y}}
\sum_{ s^{i}_V, i=1,2} \tilde F \tilde G \prod_{X \in {\cal X}} f_X
\prod_{Y \in {\cal Y}}
g_Y \exp (-{\cal H}^0_V) {\cal W} \exp (2 \sum_Y \bar \phi (Y)) \en where
${\cal
W}=e^{2f|\Lambda \backslash V|} \prod_{\alpha,i=1,2} W ( s^i_{V_\alpha})$
(${V_\alpha}$
are the connected components of $V$), and the sum over $Y$ runs over $Y
\cap (\Lambda
\backslash V) \neq \emptyset$.

Now comes the main observation: for each term of (26), decompose $V \cup
\{Y |Y \in {\cal
Y}\}$ into connected components. If {\underline A} and {\underline B} are
in different
components, then the corresponding term in (26) vanishes. Indeed, because
of the factor
$\tilde F \tilde G$, the summand is then odd under the interchange of
$s^1_i$ and $s^2_i$
for all $i$ in the component containing {\underline A} (or {\underline B}).
Here, we use
the fact that, since $L>R$, $R$ being the range of $\Phi_0$, every $X\subset V$
contributing to the sum defining ${\cal H}^0_V$, belongs to a single
connected component
$V_\alpha$.

So, for each non-vanishing term in (26), we can choose a ``connected path" $P =
(Z_i)^n_{i=1}$ where $Z_1 = {\underline A}, Z_n = {\underline B}$, each
$Z_i$ for $i \neq
1,n$ is either an ${\underline X} $ or a $Y$ and $Z_{i+1}$ is adjacent to $
Z_i$ (i.e. $d
(Z_{i}, Z_{i+1}) \leq1$), $\forall i = 1,\cdots,n-1$. So we get a bound on
(26): \be
&&|(26)| \leq Z (\Lambda)^{-2} 4 \| F \| \| G \|  \sum_P \prod_{{\underline
X} \in P} \|
f_X \| \prod_{Y \in P} \| g_Y \| \sum_{{\cal X,Y}}{^P} \sum_{ s^{i}_V}
\prod_{X \in {\cal
X}, {\underline X} \notin P} f_X \prod_{X \in {\cal X}, {\underline X} \in
P} (1+f_X)
\nonumber \\ &&\prod_{Y \in {\cal Y}\backslash P} g_Y \prod_{Y \in P}
(1+g_Y) \exp (-{\cal
H}^0_V) {\cal W} \exp(2 \sum_Y \bar \phi (Y)) \en where we used the
positivity of $f_X$, $
g_Y$, ${\cal W}$ (in particular, in order to insert the products  over
$1+f_X$, $1+g_Y$),
we bounded $\tilde F, \tilde G$ by $4 \| F \| \| G \| $, and we wrote
$\sum^P$ to denote
the sum over the pairs $({\cal X},{\cal Y})$ for which the chosen path is
$P$. Now,
observe that, by resumming the expansions,  the sum over ${\cal X,Y},
s^{i}_V$ is less
than $Z(\Lambda)^2$: we get an upper bound on $\sum^P_{{\cal X,Y}}$ by
resumming first
over all ${\cal Y}$ that are sets of $Y$'s intersecting $ \Lambda
\backslash V$, with $Y
\notin P$ and then over all ${\cal X}$ that are sets of $X$'s in $ \Lambda
$, with
${\underline X} \notin P$ (since each term in (27) is positive). This upper
bound is equal
to $Z(\Lambda)^2$. So, we finally get

\be (27) \leq 4 \| F \| \| G \| \sum_P \prod_{{\underline X} \in P} \| f_X
\| \prod_{Y \in
P} \| g_Y \| \en

Now, to get (12), we have to bound the sum over $P$. First use \be d(A,B)
\leq cL (1+
\sum_{{\underline X} \in P} d(X)) + \sum_{Y \in P} d(Y) \en for some $c<
\infty$, where
the factor $cL$ enters because we consider $Z_1 = {\underline A}$, $Z_n =
{\underline B}$,
or $Z_i=\underline X$, which are unions of $L$-cubes. We choose $m =
\frac{\gamma}{2cL}$,
and use (29) to bound $e^{md({A},{B})}$. Finally, we have to control the
sum over $P$ in
(28). Using (29) and (15,20,26,27,28), we bound $$ e^{md(A,B)} | \langle FG
\rangle_\Lambda - \langle F \rangle_\Lambda \langle G \rangle_\Lambda |
\leq 2 \| F \| \;
\| G \| \sum^\infty_{n=2} \sum_{(Z_i)^{n-1}_2} \prod^{n-1}_{i=2} h(Z_i) $$
where each
$Z_i$ is either a $\underline X$ or a $Y$, $h(Z) = \exp (\frac{\gamma
d(X)}{2}) \| f_X \|
$ or $\exp (\frac{\gamma d(Y)}{2cL}) \| g_Y \| $ and each $Z_i$ is
adjcacent to $Z_{i-1}$.
We can bound the sum over $(Z_i)$ by the (ordered) product $$\prod^{n-1}_{i=2}
(\sum_{Z_i}{^{i-1}} h (Z_i)), $$ where $\sum^{i-1}$ means that $Z_i$ is
adjacent to
$Z_{i-1}$. Now, $$ \sum_{Z_i}{^{i-1}} h (Z_i) \leq C |X| (\sum_{0 \in Z}
h(Z)) $$ if
$Z_{i-1}$ is an $\underline X$, or $$ \sum_{Z_i}{^{i-1}} h(Z_i) \leq C |Y|
(\sum_{0 \in Z}
h(Z)), $$ if $Z_{i-1}$ is a $Y$, since the number of translates of $Z_i$,
$0 \in Z_i$ that
are adjacent to $Z_{i-1}$ (with $Z_{i-1}={\underline X}$ or $Y$) is bounded
by a constant
times $|{ X}|$ or $|Y|$, respectively. We then use the bounds \be
\sum_{0\in {\underline
X}} |{ X}| \exp(\frac{\gamma d(X)}{2}) \| f_X \| \leq C L^d \epsilon, \en
and \be
\sum_{0\in {Y}} |{Y}| \exp(\frac{\gamma d(Y)}{2cL}) \| g_Y \| \leq C L^d
\epsilon, \en
which follow from (19,25) (in (30), we use the fact that the sum over $0\in
{\underline
X}$ is bounded by $C L^d$ times the sum over $0\in { X}$, and $|X| \leq C
d(X)^d$ while in
(31), we use $|Y|\leq C d(Y)^d \leq C (2cL)^d (\frac{d(Y)}{2cL})^d$).
Finally, for
$\epsilon$ small enough, we bound (28) by a convergent geometric series: $$
e^{md(A,B)} |
\langle FG \rangle_\Lambda - \langle F \rangle_\Lambda \langle G
\rangle_\Lambda | \leq 2
\| F \| \; \| G \| \sum^\infty_{n=0} (CL^d \epsilon)^n $$

To prove the existence of the $\Lambda \uparrow {\bf Z}^d$ limit (in the
sense of finite
subsets ordered by inclusion) and the uniqueness of the Gibbs state,
consider $F$ as in
(12) and two boundary conditions $s\in \Omega_{\Lambda^c}, s' \in
\Omega_{\Lambda^{'c}}$.
Compare \be \langle F \rangle_\Lambda (s) -\langle F \rangle_{\Lambda'} (s') =
\sum_{s^1_\Lambda,{s}^2_{\Lambda'}} \tilde F \nu ({s}^1_\Lambda | {s}) \nu
({s}^2_{\Lambda'} |{s}') \en with $\Lambda$, $\Lambda'$ such that $A
\subset \Lambda \cap
\Lambda'$, and do the same expansion as above. The only terms that do not
vanish are such
that the connected component containing $\underline{A}$ intersects
$\partial (\Lambda \cap
\Lambda')$, because $\tilde F$ is antisymmetric under the ${s}^1
\leftrightarrow {s}^2$
interchange. Hence, for each non-zero term, we may choose a path $P$ connecting
$\underline{A}$ to $\partial (\Lambda \cap \Lambda')$. By the arguments
given above, the
sum over those paths is bounded by $\| F \| \exp (-m d(A,\partial (\Lambda \cap
\Lambda')))$, and (32) goes to zero as $\Lambda, \Lambda' \uparrow {\bf
Z}^d$.\hfill$
\makebox[0mm]{\raisebox{0.5mm}[0mm][0mm]{\hspace*{5.6mm}$\sqcap$}}$ $
\sqcup$ \vspace*{3mm}

In the next sections, we shall need a simple extension of Theorem 1 to a
situation where
the interactions depend (weakly) on $\Lambda$. Assume that, for each finite
$\Lambda
\subset {\bf Z}^d$, $\Phi_\Lambda = \Phi^0_\Lambda + \Phi^1_\Lambda$ where
$\Phi^0_\Lambda$, $\Phi^1_\Lambda$ satisfy the hypotheses of Theorem 1
uniformly in
$\Lambda$. Assume also that \be \| \Phi_{\Lambda X} - \Phi_{\Lambda' X} \|
\leq \epsilon
e^{-\gamma d(X,\partial \Lambda)} \en for $X \subset \Lambda \subset
\Lambda'$, so that
$\Phi_X = \lim_{\Lambda \uparrow {\bf Z}^d} \Phi_{\Lambda X}$ exists. Then
we have

\begin{Th} Under the above assumptions, the conclusions of Theorem 1 hold;
moreover,
$\lim_{\Lambda \uparrow {\bf Z}^d} \nu_\Lambda = \mu$, where $\nu _\Lambda$
is defined by
(2-4) with $\Phi_X$ replaced by $\Phi_{\Lambda X}$ and the sum in (2) is
restricted to $X
\subset \Lambda$. \end{Th}

\vspace*{3mm} \par\noindent {\bf Proof} We study the limit $\Lambda
\uparrow {\bf Z}^d$ of
$\nu_\Lambda$. Let $\tilde \Lambda \subset \Lambda$ denote the largest cube
containing the
origin in ${\bf Z}^d$ such that $| \tilde \Lambda | \leq \frac{\gamma}{4}
d(\Lambda)$ (so
that $d(\tilde \Lambda) \to \infty$ if $d(\Lambda) \to \infty$).  Write the
Hamiltonian $$
{\cal H} (s_\Lambda) = \sum_{X \subset \tilde \Lambda} \Phi_X +
\Phi_{\tilde \Lambda} +
\sum_{X \not \subset \tilde \Lambda} \Phi_{\Lambda X} $$ where
$\Phi_{\tilde \Lambda} =
\sum_{X \subset \tilde \Lambda} \Phi_{\Lambda X} - \Phi_X$ can be regarded as
an
interaction coupling all the variables in ${\tilde \Lambda}$ and $\Phi_X =
\lim_{\Lambda
\uparrow {\bf Z}^d} \Phi_{\Lambda X}$. By (33), we have the trivial bound:
\bea \|
\Phi_{\tilde \Lambda} \| \leq \sum_{X \subset \tilde \Lambda} \|
\Phi_{\Lambda X} - \Phi_{
X} \| \leq \epsilon 2^{|\tilde \Lambda|} e^{-\gamma d (\tilde \Lambda,
\partial \Lambda)}
\leq C\epsilon e^{-\frac{\gamma}{3} d(\tilde \Lambda)}. \eea

So, $\Phi_{\tilde \Lambda}$ satisfies an estimate similar to the one of
$\Phi^1$ (with a
smaller $\gamma$). If we introduce this representation of ${\cal H}$ and expand
$\Phi_{\tilde \Lambda}$, considering it as a part of $\Phi^1$, we see that
the only
non-zero term must contain a path connecting $A$ to $\partial \tilde
\Lambda$. Since
$d(\tilde \Lambda) \to \infty$ when $\Lambda \uparrow {\bf Z}^d$, the
existence of the
limit for $\nu_\Lambda$ follows. The rest is then as in the proof of
Theorem 1. \hfill$
\makebox[0mm]{\raisebox{0.5mm}[0mm][0mm]{\hspace*{5.6mm}$\sqcap$}}$ $ \sqcup$

\vspace*{5mm} \par\noindent {\bf Remark 1.} We used in an essential way the
fact that
$\Phi$ is real when we used the positivity of $f_X,g_X,{\cal W}$. This can
always be
arranged for $\Phi$ real by adding a constant but not,  of course, for
complex $\Phi$.
This ``trick" of taking advantage of the positivity of the interactions was
used e.g. in
\cite{BK,VKP,Fi,Ma}. The main interest of our method is that it yields
uniqueness of the
Gibbs state and exponential decay of correlations without using much
combinatorics.

\vspace*{5mm} \par\noindent {\bf Remark 2.} Here we first had $L$
determined by the
$\Phi_0$ part of the interaction and then we chose $\epsilon$ small enough.
However, in
applications such as those of the next sections, the splitting of the
interaction into
$\Phi_0$ and $\Phi_1$ is somewhat arbitrary, and $L$ and $\epsilon$ are not
independent.
All we need is that $L^d \epsilon$ be small enough, see (30,31).

\vspace*{5mm} \par\noindent {\bf Remark 3.} The exponential decay (12)
implies the decay of
all the truncated correlation functions and, therefore, the fact the free
energy and the
correlation functions are $C^\infty$ (in the same sense as analyticity was
defined above);
see e.g. \cite{S}, Sect. II 12. Of course, the bounds obtained in this way
on the truncated
correlation functions contain factorials that prevent us from drawing any
conclusions
about their analyticity.

\vspace*{5mm} \par\noindent {\bf Remark 4.} Since we used $\| \Phi^1\|_2
\leq \epsilon$
only in the bounds (19,30), we can easily extend our results in several
directions: First,
one recovers uniqueness of the Gibbs state if we assume only $\| \Phi^1
\|_1 \leq
\epsilon$, and $\Phi^0$ as before. Instead of (19), we have \be \sum_{0 \in
X} |X| \| f_X
\| \leq {\cal O} (\epsilon) \en and (30) holds without the factor
$\exp(\frac{\gamma}{2}d(X))$.  Taking $\Phi^0 = 0$, this yields a simple
proof of
Dobrushin's uniqueness theorem. Using (30,31) (without
$\exp(\frac{\gamma}{2}d(X))$) we
can also bound \be \sum_{i \in {\bf Z}^d}| \langle F \tau_i G \rangle -
\langle F \rangle
\langle G \rangle | < \infty \en (just sum over $Z_n = \tau_i {\underline
B}$). Therefore,
one recovers the result of Gross \cite{Gr2} on the $C^2$ property of the
free energy.

We may also consider potentials decreasing with a power law: introduce the
norm \be \|
\Phi \|_4 = \sum_{0 \in X} |X| e^{\gamma \ln d(X)} \| \Phi_X \| \en Then,
if $\| \Phi^1
\|_4$ is small, we get an upper bound $d(A,B)^{-\gamma}$ in (12). Of
course, here we need
the factor $|X|$ in (36) to control the sum (30).

Finally, let us remark that we do not use the translation invariance of
$\Phi$ except when
we speak of the free energy, of analyticity or of differentiability. Hence,
uniqueness of
the Gibbs state and the decay of correlation (12) hold for arbitrary
interactions on
arbitrary lattices with $$ \sup_i \sum_{i \in X} |X| \| \Phi^1_X \| \;\;
\mbox{or} \;\;
\sup_i \sum_{i \in X} e^{\gamma d(X)} \| \Phi^1_X \| $$ small enough.

\setcounter{equation}{0} \section{Coupled map lattices.}

We consider the following class of dynamical systems. The phase space
${\cal M} =
(S^1)^{{\bf Z}^d}$ i.e. ${\cal M}$ is the set of maps ${\bf z} =(z_j)_{j\in
{\bf Z}^d}$
from ${\bf Z}^d$ to the circle. ${\cal M}$ inherits its topology and Borel
$\sigma$-algebra ${\cal B}$ from $S^1$ : it is a compact metric space and
we let $m$
denote the product of Lebesgue measures on the $S^1$-factors. ${\cal M}$
carries a natural
${\bf Z}^d$-action, denoted by $\tau_i$, $i \in {\bf Z}^d$.

To describe the dynamics, we first fix a map $F: S^1 \to S^1$. We take $F$
to be an
expanding, orientation preserving $C^{1+\delta}$ map with $\delta >0$. We
describe $F$ in
terms of its lift to ${\bf R}$, denoted by $f$ and chosen, say, with $f(0)
\in [0,1[$.  We
assume that \be f'(x) > \lambda^{-1} \en where $\lambda < 1$. Note that
there exists an
integer $k > 1$ such that \be f(x+1) = f(x) + k \;\;\;\; \forall x \in {\bf
R} \en We let
${\cal F}: {\cal M} \to {\cal M}$ denote the Cartesian product ${\cal F} =
{\rm X}_{i \in
{\bf Z}^{d}} F_i$ where $F_i$ is a copy of $F$. ${\cal F}$ is called the
uncoupled map.

The second ingredient in the dynamics is given by the coupling map $A:
{\cal M} \to {\cal
M}$. This is taken to be a small perturbation of the identity in the
following sense. Let
$A_i$ be the projection of $A$ on the i$^{\mbox{th}}$ factor. We assume
that $A_i$ is
$C^1$ in each $z_j$ and satisfies (we use the parametization $z_j = e^{2\pi
i x_j}$ on
$S^1$, $x_j \in [0,1[$  and we let $a_j$ denote the lift of $A_j: A_j =
e^{2 \pi i a_j}$):
\be \| \frac{\partial a_i}{\partial x_j} - \delta_{ij} \|_\infty \leq
\epsilon e^{-\beta
|i-j|} \en together with a H\"older condition \be \Bigg | \frac{\partial
a_i}{\partial
x_j} ({\bf x}) - \frac{\partial a_i}{\partial x_j} ({\bf y}) \Bigg | \leq
\epsilon \sum_k
e^{-\beta(|i-j|+|i-k|)} |x_k - y_k|^\delta \en for some $\epsilon, \beta,
\delta > 0$,
with ${\bf x}= (x_i)_{i\in {\bf Z}^d}$.

Furthermore, we suppose that $A$ is ${\bf Z}^d$-invariant: \be A= \tau_i
\circ A \circ
\tau^{-1}_i \;\; \forall i \in {\bf Z}^d, \en although this is not
essential, see Remark 2
in \cite{BK2} and Remark 4 at the end of Section 2.

\vs{3mm}

\no {\bf Example.} An example often considered (see e.g. \cite{BK2}) is the
coupling map $$
a_j({\bf x}) = x_j + 2 \pi i \epsilon \sum_k g_{|j-k|} (x_j , x_k) $$ where
$g$ is a
periodic $ C^{1+\delta}$ function in both variables, with exponential
falloff in $|j-k|$
as in (4). More general examples of such $A's$ can be found in \cite{BK2}
(where, however,
we restricted ourselves to analytic maps).

\vs{3mm}

The coupled map $T: {\cal M} \to {\cal M}$ is now defined by $$T = A \circ
{\cal F}.$$ We
are looking for ``natural" $T$-invariant measures on ${\cal M}$. For this,
write, for
$\Lambda \subset {\bf Z}^d$, ${\cal M}_\Lambda = (S^1)^\Lambda$, and let
$m_\Lambda$ be
the product of Lebesgue measures. Let ${\cal B}_\Lambda \subset {\cal B}$ be
the
$\sigma$-algebra of subsets of ${\cal M}$ generated by the Borel sets of ${\cal
M}_\Lambda$ (identified in the obvious way with subsets of ${\cal M}$).

\vspace*{5mm} \par\noindent {\bf Definition 1} {\em A Borel probability
measure $\mu$ on
${\cal B}$ is a SRB measure if \begin{enumerate} \item[(a)] $\mu$ is
$T$-invariant
\item[(b)] The restriction $\mu_\Lambda$ of $\mu$ to ${\cal B}_\Lambda$ is
absolutely
continuous with respect to  $m_\Lambda$ for all $\Lambda \subset {\bf Z}^d$
finite.
\end{enumerate}}

\vspace*{4mm} \par\noindent {\bf Remark} This is a natural extension of the
notion of SRB
measure to infinite dimensions, since each $S^1$ factor can be regarded as
an unstable
direction.

\vspace*{4mm}

It is convenient to introduce the space of H\"older continuous functions
${\cal C}^\delta
({\cal M}_\Lambda)$ for $\Lambda \subset {\bf Z}^d, | \Lambda | < \infty$
equipped with
the norm \bea \| G \|_\delta = \| G \|_\infty + \sup_{{\bf x,y}}
\frac{|G({\bf x}) - G
({\bf y})|}{\sum_{i \in \Lambda} |x_i - y_i |^\delta} \eea We also write
${\cal C} ({\cal
M}_\Lambda)$ to denote the continuous functions on ${\cal M}_\Lambda$ with
the sup norm.

\vspace*{4mm} \par\noindent {\bf Definition 2} {\em A Borel probability
measure on ${\cal
B}$ is regular if its restriction $\mu_\Lambda$ to ${\cal B}_\Lambda$ is
absolutely
continuous with respect to $m_\Lambda$ for all $\Lambda \subset {\bf Z}^d$
finite and if
$\log h_\Lambda \in {\cal C}^\delta ({\cal M}_\Lambda)$, where $h_\Lambda =
\frac{d\mu_\Lambda}{dm_\Lambda}$: \be | \log \frac{h_\Lambda ({\bf
x})}{h_\Lambda ({\bf
y})} | \leq C \sum_{i \in \Lambda} | x_i - y_i |^\delta \en }

\vspace*{5mm} \par\noindent {\bf Remark} A similar condition was introduced
by Volevich
\cite{V}. If $\mu$ is a Gibbs state on ${\cal M}$ for some translation
invariant
interaction such that $\sum_{0\in X} \| \Phi_X \| < \infty$ and \be |
\Phi_X ({\bf x}) -
\Phi_X ({\bf y}) | \leq C (X) \sum_{i \in X} | x_i - y_i |^\delta \en with
\be \sum_{0\in
X} C(X) < \infty, \en then $\mu$ is regular. This follows easily from the
DLR equations:
\be h_\Lambda ({\bf x}) = \int \nu ({\bf x} | {\bf z}) d\mu ({\bf z}) \en
where $\nu ({\bf
x} | {\bf z})$ is defined as in (2.3). To get (6), using (9), by Jensen's
inequality it is
enough to bound \bea &&\sup_{{\bf z}} | \log \frac{\nu ({\bf x} | {\bf
z})}{\nu ({\bf y} |
{\bf z})} |= \sup_{{\bf z}} | {\cal H} ({\bf x} | {\bf z}) - {\cal H} ({\bf
y} | {\bf z})
| \\ && \leq \sup_{{\bf z}} \sum_{X \cap \Lambda \neq \phi}  | \Phi_X ({\bf
x} \vee {\bf
z}) - \Phi_X ({\bf y} \vee {\bf z})  | \eea which, using (7, 8), is bounded
by the RHS of
(6).

\vspace{2mm}

We will prove \begin{Th} Let $F$ and $A$ satisfy the assumptions given
above. Then there
exists $\epsilon_0 > 0$ such that, for $\epsilon < \epsilon_0$, $ T$ has an
SRB measure
$\mu$. Furthermore, $\mu$ is invariant and exponentially mixing under the
space-time
translations generated by $\tau_i$ and $T$: there exists $m>0$,
$C<\infty$, such that, $\forall B,D \subset {\bf Z}^d, |B|,|D| < \infty$ and
$\forall G \in L^{\infty}({\cal M}_B), \forall H \in {\cal C}^\delta ({\cal
M}_D)$,
\be |\int G \circ T^n Hd\mu - \int G d \mu \int H d \mu| \leq C \| G
\|_\infty \| H \|_\delta \min (|B|,|D|) e^{-m(n+d(B,D))}, \en
 where $d(B,D)$
is the distance between $B$ and $D$.

Finally, for any regular measure $\nu$, $T^{*N} \nu \to \mu$ weakly, as $N
\to \infty$,
$T^{*}$ being the transpose of $T$. \end{Th}

\no {\bf Remark.} We shall construct an expansion for the Perron-Frobenius
operator of $T$,
which will be quite similar to the expansion of the previous section, for
$\Phi$ of the
form (2.9), with $\Phi^0$ one-dimensional and $\| \Phi^1 \|_2 $ small.
Bunimovich and
Sinai \cite{BS} have reduced the problem (for a one dimensional lattice of
maps, and a
slightly different model) to a spin system with exponentially decaying
interactions of the
same form. We shall rederive this estimate below (see Proposition 2). However,
they did not discuss in detail the resulting statistical mechanical system (the
reference to \cite{DM1} is slightly misleading, since the latter authors
assume $\| \Phi^1 \|_3$ small, which probably does not hold in general here).
The work of Bunimovich and Sinai was extended by Volevich \cite{V,V2,V3},
Pesin and Sinai \cite{PS}, and by Jiang \cite{J}. Our results are quite
close to those of \cite{V3}, where lattices of dimension $d>1$ are
considered. In that paper, the statistical mechanics is based on an
extension to infinite range interactions of results of
Dobrushin and Pecherski \cite{DP}.

An extension of \cite{BS} is also claimed
in \cite{GR,CR}. However, there the treatment of the statistical mechanics part
of the problem is wrong. The authors essentially claim that uniqueness of the
Gibbs state holds for any $ \Phi $, with $\| \Phi \|_2 $ finite. Although this
is true in one dimension, it certainly does not hold in higher dimensions. The
Ising model, or any finite range model of this type that undergoes a phase
transition, provide counterexamples. Moreover, the ``proof" would also yield
analyticity, which does not hold even under the correct assumption that $\Phi$
is of the form (2.9), with $ \Phi^0 $ one dimensional and $\| \Phi^1 \|_2$
small, because of the counterexamples in \cite{DM2}.

 Finally, in \cite{BK2},
we constructed a convergent expansion for the Perron-Frobenius operator, but
we needed stronger assumptions on the coupled map lattice (analyticity instead
of smoothness). These assumptions, in the present context, would imply that
$\| \Phi^1 \|_3$ is small and that we could use the usual cluster expansion.

\setcounter{equation}{0} \section{SRB measure from a Gibbs measure}

The SRB measure $\mu$ is constructed using the Perron-Frobenius  operator
for $T$. We
first introduce a finite volume cutoff version of $T$, $T_\Lambda : {\cal
M}_\Lambda \to
{\cal M}_\Lambda$,  $\Lambda \subset {\bf Z}^d$, $|\Lambda| < \infty$. For
definiteness,
we let $\Lambda$ be a cube centered at 0. Set $T_\Lambda = A_\Lambda \circ
{\cal
F}_\Lambda$ with ${\cal F}_\Lambda= {\times}_{i \in \Lambda} F_i$ and \be
A_\Lambda =
R_\Lambda \circ A \circ {\cal C}_\Lambda \en where ${\cal C}_\Lambda :
{\cal M}_\Lambda
\to {\cal M}$ extends ${\bf z} \in {\cal M}_\Lambda$ to ${\cal M}$ by $$
{\cal C}_\Lambda
({\bf z})_i = \left\{ \begin{array}{ccc} z_i && i \in \Lambda \\ 1 && i
\not \in \Lambda
\end{array} \right. $$ (here $z_i, 1 \in S^1$, and the choice of $1$ is
arbitrary), and
$R_\Lambda : {\cal M} \to {\cal M}_\Lambda$ is the restriction. It is more
convenient to
work in the covering space ${\bf R}^\Lambda$ of ${\cal M}_\Lambda$ and to
introduce the
corresponding lifts $a_\Lambda, f_\Lambda, t_\Lambda$ as maps from ${\bf
R}^\Lambda$ to
itself. Our assumptions for $A$ then imply the following properties for
$a_\Lambda$.

\begin{Lem} The maps $a_\Lambda$ are $C^{1+\delta}$ diffeomeorphisms with
the following
properties: there exist $ \beta > 0$, $ C < \infty$ , independent of $\Lambda$,
such that
$\forall {\bf x} \in{\bf R}^\Lambda$ \begin{enumerate} \item[a)] $a_\Lambda
({\bf x} +
{\bf n}) = a_\Lambda ({\bf x}) + {\bf n} \;\; \forall {\bf n} \in {\bf
Z}^\Lambda$
\item[b)] The derivative of $a_\Lambda$ can be written as $D a_\Lambda
({\bf x}) = I +
\epsilon_\Lambda ({\bf x})$, where $I$ is the unit matrix, and
$\epsilon_\Lambda$ is
periodic function satisfying \be | \epsilon_\Lambda ({\bf x})_{ij} | \leq C
\epsilon
e^{-\beta | i-j|} \en \be | \epsilon_\Lambda ({\bf x})_{ij} -
\epsilon_\Lambda ({\bf
y})_{ij} | \leq C\epsilon \sum_{k \in \Lambda}
e^{-\beta(|i-j|+|i-k|)}|x_k-y_k|^\delta \en
\item[c)] $\epsilon_\Lambda ({\bf x})_{ij} = \epsilon_{\Lambda'} ({\bf
x}\vee  {\bf
1}_{\Lambda' \backslash \Lambda})_{ij}$ for $i,j \in \Lambda \subset
\Lambda'$. \item[d)]
Let $\Lambda \subset \Lambda'$, $R_\Lambda {\bf y} = {\bf x}$, ${\bf y} \in
{\bf
R}^{\Lambda'}$ and $i,j \in \Lambda$. Then \be | \epsilon_\Lambda ({\bf
x})_{ij} -
\epsilon_{\Lambda'} ({\bf y})_{ij} | \leq C \epsilon e^{-\beta |i-j|} e^{-\beta
d(i,\partial \Lambda)} \en where $d$ denotes the distance. \end{enumerate}
Furthermore
$a_\Lambda^{-1}$ satisfies a)-d) also, with possibly different $ \beta$,$
C$. \end{Lem}

\vspace*{5mm} \par\noindent {\bf Proof} (2) and (3) are just a restatement
of (3.3), (3.4)
(we can take $C=1$ here). By definition of a lift, $a_\Lambda ({\bf x} +
{\bf n}) =
a_\Lambda ({\bf x}) + {\bf m(n)}$ for some ${\bf m} ({\bf n}) \in {\bf
Z}^\Lambda$. Write
$a_\Lambda = id + b$ where $id$ is the identity. Order points of $\Lambda$
in an arbitrary
way such that $\Lambda = \{ i_\ell | \ell = 1, \cdots, |\Lambda |\}$. Then,
given ${\bf
x}, {\bf y} \in {\bf R}^\Lambda$, define ${\bf w}^{(n)}$ by letting let
$w^{(n)}_{i_\ell}$
equal $x_{i_\ell}$ for $\ell \geq n$ and $y_{i_\ell}$ for $\ell < n$. Then,
$$ b_i ({\bf
x}) - b_i ({\bf y}) = \sum^{|\Lambda |}_{n=1} b_i ({\bf w}^{(n)}) - b_i
({\bf w}^{(n+1)})
$$ and, from (2), \BE |b_i ({\bf w}^{(n)}) - b_i ({\bf w}^{(n+1)}) | &\leq&
\int^1_0 dt |
\epsilon_{ii_n} ((1-t) {\bf w}^{(n)} + t {\bf w}^{(n+1)}) | | x_{i_n} -
y_{i_n} |
\nonumber \\ &\leq& C\epsilon e^{-\beta | i-i_n|} | x_{i_n} - y_{i_n} |.
\EN Hence, \be |
b_i ({\bf x}) - b_i ({\bf y}) | \leq C\epsilon \sum_{j \in \Lambda}
e^{-\beta |i-j|} |x_j
- y_j |. \en Let $| n_j | \leq 1$. Then, for $\epsilon$ small, ${\bf m}
({\bf n}) = {\bf
n}$ and thus by iteration a) holds for all ${\bf n}$. Point c) follows
from the
definition (1) and d) follows from c) and (3).

It is now straightforward to show that $a^{-1}_\Lambda$ exists and
satisfies a)-d) too,
(possibly with different $\beta$, $C$): Writing $D a^{-1}_\Lambda = 1 + \tilde
\epsilon_\Lambda = (1 + \epsilon_\Lambda)^{-1}$ and using (2) with $C=1$,
we get: \be |
\tilde \epsilon_\Lambda ({\bf x})_{ij} | \leq \sum^\infty_{n=1} |
\epsilon^n_\Lambda ({\bf
x})_{ij} |  \leq \sum^\infty_{n=1} \epsilon^n \sum_{i_2,\cdots,i_{n-1}}
\prod^{n-1}_{\ell
= 1} e^{-\beta | i_\ell - i_{\ell +1}|} \nonumber \en where $i_1 = i$, $i_n
= j $. This is
readily bounded by $ \epsilon e^{- \tilde \beta |i-j|} $ for any $\tilde
\beta < \beta$
and $C = C(\tilde \beta) < \infty$. This proves (2) for $a^{-1}_\Lambda$.
To prove (3) for
$a^{-1}_\Lambda$, we use (3) for $\epsilon_\Lambda$ and \be |
\epsilon^n_\Lambda ({\bf
x})_{ij} - \epsilon^n_\Lambda ({\bf y})_{ij} | \leq \epsilon^n
\sum_{i_2,\cdots,i_{n-1}}
\sum^n_{\ell = 1} \sum_k e^{-\beta | i_\ell - k |} \prod^{n-1}_{\ell=1}
e^{-\beta
|i_\ell-i_{\ell+1}|} |x_k - y_k |^\delta \nonumber \en with $i_1 = i, i_n =
j$, which is
bounded by \be \leq (C \epsilon)^n \sum_k e^{- \tilde \beta (|i-j| +
|i-k|)} |x_k - y_k
|^\delta \en with $\tilde \beta < \frac{\beta}{2}$ and $C$ as above. The
proof of (4) is
similar. \hfill$
\makebox[0mm]{\raisebox{0.5mm}[0mm][0mm]{\hspace*{5.6mm}$\sqcap$}}$ $
\sqcup$

\vspace*{5mm}

The Perron-Frobenius operator $P_\Lambda$ for $T_\Lambda$ is defined as
usual by \be \int
G \circ T_\Lambda H d m_\Lambda = \int G P_\Lambda H dm_\Lambda \en for $G
\in L^\infty
({\cal M}_\Lambda), H \in L^1 ({\cal M}_\Lambda)$. Let us work in the covering
space ${\bf
R}^\Lambda$ and replace $G, H$ by periodic functions denoted $g, h: g ({\bf
x} + {\bf n})
= g ({\bf x})$, $ \forall {\bf n} \in {\bf Z}^\Lambda$. Insert $t_\Lambda =
a_\Lambda
\circ f_\Lambda$ in (8) to get (with $C_\Lambda = [0,1 ]^\Lambda)$ \be
\int_{C_\Lambda} g
\circ t_\Lambda h d {\bf x} = \int_{C_\Lambda} g (a_\Lambda ({\bf x}))
P^\circ_\Lambda h
({\bf x}) d{\bf x} \en where $P^\circ_\Lambda$ is the Perron-Frobenius
operator for
$f_\Lambda$ i.e. $$ P^\circ_\Lambda h ({\bf x}) = \sum_{{\bf s}}
\frac{h(f^{-1}_\Lambda
({\bf x} + {\bf s}))}{\prod_{i \in \Lambda} f' (f^{-1} (x_i + s_i))} $$
where ${\bf s} \in
\{0,\cdots,k-1\}^\Lambda$ (and $k$ was introduced in (3.2)). Note that
$P^\circ_\Lambda$
maps periodic functions into periodic functions because the sum is periodic
even if the
summands are not: indeed, (3.2) implies that $f^{-1} (x+1+k-1) = f^{-1}
(x+k) = f^{-1} (x)
+ 1$ (so that, if we add $1$ to $x_i$, it amounts to a cyclic permutation
of $s_i$) and
that $f'$ is periodic.

Since, for a periodic function $k$ $$ \int_{a_\Lambda (C_\Lambda)} k({\bf
x}) d{\bf x}  =
\int_{C_\Lambda} k({\bf x}) d{\bf x} $$ (both $C_\Lambda$ and $a_\Lambda
(C_\Lambda)$ are,
up to their boundaries, fundamental domains for the action of ${\bf
Z}^\Lambda$ on ${\bf
R}^\Lambda$), we obtain the formula \be P_\Lambda h({\bf x}) = \det D
a^{-1}_\Lambda ({\bf
x}) \sum_{{\bf s}} \frac{h(\psi_{\Lambda{\bf s}}({\bf x}))} {\prod_{i \in
\Lambda} f'
(\psi_{\Lambda{\bf s}} ({\bf x})_i)} \en where we defined \be
\psi_{\Lambda{\bf s}} ({\bf
x})_i = f^{-1} (a^{-1}_\Lambda ({\bf x})_i + s_i). \en and (with a slight
abuse of
notations) we write $P_\Lambda$ for the operator acting on periodic
functions induced by
the Perron-Frobenius operator defined by (8). Note that, by Lemma 4.1.a,
$a^{-1}_\Lambda
({\bf x} + {\bf n}) = a^{-1}_\Lambda ({\bf x}) + {\bf n}$ and, so,
$P_\Lambda$ maps also
periodic functions into periodic functions.

The invariant measure $\mu$ is constructed as a weak limit of the measures
$T^{\ast
N}_\Lambda m_\Lambda$, with $T^\ast_\Lambda$ being the transpose of
$T_\Lambda$, as $N \to
\infty$ and $\Lambda \uparrow {\bf Z}^d$. These in turn are given by the
Perron-Frobenius
operator $$ T^{\ast N}_\Lambda m_\Lambda = (P^N_\Lambda 1) m_\Lambda $$ by
putting $H = 1$
in (8). $P^N_\Lambda 1$ has a direct statistical mechanical interpretation
which we now
derive.

First, iterating (10), we get \be (P^N_\Lambda 1) ({\bf x}) = \sum_{{\bf
s}_1, \cdots {\bf
s}_N} \prod^N_{t=1} \Bigg[ \det D a^{-1}_\Lambda (\psi_{{\bf s}_{t-1}}
\circ \cdots \circ
\psi_{{\bf s}_1} ({\bf x})) \prod_{i \in \Lambda} (f' (\psi_{{\bf s}_t}
\circ \cdots \circ
\psi_{{\bf s}_1} ({\bf x})_{i}))^{-1} \Bigg] \en where, for $t=1$, the
argument of $D
a^{-1}_\Lambda$ is ${\bf x}$, and we write $\psi_{\bf s}$ for
$\psi_{\Lambda{\bf s}}$.
{}From now on,  we shall consider ${\bf x}\in C_\Lambda$.

Next, we introduce a convenient notation: ${\bf x}\in C_\Lambda$  and ${\bf
s}_1, \cdots,
{\bf s}_N$ in (12) collectively define a configuration on a ``space-time"
lattice
$\{0,\cdots,N\} \times \Lambda$. Thus, let ${\bf Z}^{d+1}_+ = {\bf Z}_+
\times {\bf Z}^d,
{\bf Z}_+$ being the non-negative integers. To any subset $X \subset {\bf
Z}^{d+1}_+$
associate the configuration space $\Omega_X = \times_{\alpha \in X}
\Omega_\alpha$ where,
for $\alpha = (t,i)$, $\Omega_\alpha$ equals [0,1] if $t=0$, and equals
$\{0,\cdots,k-1\}$
if $t>0$. We could use the existence of a Markov partition for $T_\Lambda$
to write ${\bf
x}$ as a symbol sequence, as is usually done, e.g. in \cite{BS}, but we
shall not need
this representation.

Let $\Lambda_N = \{ 0,\cdots,N\} \times \Lambda$ and ${ s} = ({\bf x},{\bf
s}_1,
\cdots,{\bf s}_N) \in \Omega_{\Lambda_N}$. We shall  write ${\bf x}$,${\bf
s}$, as before,
for elements of $\Omega_{\Lambda}$, $\Lambda \subset {\bf Z}^{d}$, and use
${ s}$ to
denote elements of $\Omega_{\Lambda}$, $\Lambda \subset {\bf Z}^{d+1}_+$
.Then (12) reads
\be (P^N_\Lambda 1) ({\bf x}) = \sum_{{\bf s}_1 \cdots {\bf s}_N} e^{- {\cal
H}_{\Lambda_N}({ s})} \en with $e^{-{\cal H}_{\Lambda_N}}$ being the
summand in (12).

The final step is to write ${\cal H}_{\Lambda_N}$ in terms of potentials.
First, write $D
a^{-1}_\Lambda = 1 + \tilde \epsilon_\Lambda$, where, by Lemma 1, $\tilde
\epsilon_\Lambda$ satisfies (2)-(4). Then expand $$ \det D a^{-1}_\Lambda =
\det (1+
\tilde \epsilon_\Lambda) = \exp [- \sum_{i \in \Lambda} v_{\Lambda i}] $$
where \be
v_{\Lambda i} ({\bf x}) = \sum^\infty_{n=1} \frac{(-1)^n}{n} (\tilde
\epsilon_\Lambda
({\bf x})^n)_{ii} \en

\begin{Lem} There exist $\epsilon_1 > 0$, $C<\infty$  such that, for $\epsilon
<
\epsilon_1$, $$ \| v_{\Lambda i} \|_\infty \leq C \epsilon $$ and $$ |
v_{\Lambda i} ({\bf
x}) - v_{\Lambda i} ({\bf y}) | \leq C \epsilon \sum_{k \in \Lambda}
e^{-\frac{\beta}{2}|i-k|} |x_k - y_k |^\delta $$ and, for $\Lambda \subset
\Lambda'$,
${\bf x} = R_\Lambda {\bf y}$, $ i \in \Lambda$ $$ | v_{\Lambda i} ({\bf
x}) - v_{\Lambda'
i} ({\bf y}) | \leq C \epsilon e^{-\frac{\beta}{2} d (i, \partial \Lambda)}
$$ uniformly
in $\Lambda, \Lambda'$. \end{Lem}

\par\noindent {\bf Proof} Straightforward, using (2)-(4), and the bound
(7), for $\tilde
\epsilon_\Lambda$. \hfill$
\makebox[0mm]{\raisebox{0.5mm}[0mm][0mm]{\hspace*{5.6mm}$\sqcap$}}$ $ \sqcup$

\vspace*{5mm}

Given $\alpha = (t,i) \in \Lambda_N$ we set \be V_{\Lambda \alpha} ({ s}) =
v_{\Lambda i}
(\psi_{{\bf s}_{t-1}} \circ \cdots \circ \psi_{{\bf s}_1} ({\bf x})) + \log f'
((\psi_{{\bf s}_t} \circ \cdots \circ \psi_{{\bf s}_1} ({\bf x}))_i) \en
and hence \be
{\cal H}_{\Lambda_N} ({ s}) = \sum_{\alpha \in \Lambda_N} V_{\Lambda
\alpha} ({ s}) \en
where $V_{\Lambda \alpha}$ depends on ${ s} |_{\Lambda_N}$. The
representation (16) still
needs to be localized in order to express ${\cal H}$ in terms of
potentials. We use the
notation of Section 2: given ${ s} \in \Omega_X$, ${ s'} \in \Omega_Y$ for
$X \cap Y =
\emptyset$ we denote by ${ s} \vee { s'}$ the corrresponding configuration
in $X \cup Y$,
and, if $Z \subset X$, we let ${ s}_Z = { s}|_Z$. Also, let ${ 0}_X \in
\Omega_X$ be $({
0}_X)_\alpha = 0$, $\forall \alpha \in X$ (note that $0$ belongs to
$\Omega_\alpha$ for
any $ \alpha$).

Now, given $t$, choose an arbitrary ordering of the points of $\{0,\dots
,t\}\times {\bf Z}^{d}=\{ \alpha_\ell | \ell \in {\bf N} \}$ in such a way
that $| \alpha_\ell |$ is a non-decreasing function of $\ell$ where, for
$\alpha = (t',j)$ \be | \alpha | = \frac{1}{M} |t'| + \sum^d_{k=1} | j_k | .
\en $M$ will be taken sufficiently large below (see (25)).

Consider $V_{\Lambda \alpha}$
in (16). Let $Y^m_\alpha = \{ \alpha - \alpha_\ell | 0 \leq
\ell \leq m \}$, and set, for ${ s} \in \Omega_{Y^m_\alpha}$
\be \Phi^\Lambda_{Y^m_\alpha}
({ s}) = V_{\Lambda \alpha} ({ s} _{Y^m_\alpha}
\vee { 0}_{(Y^m_\alpha)^c}) - V_{\Lambda \alpha} ({ s}_{Y^{m-1}_\alpha}
\vee {
0}_{(Y^{m-1}_\alpha)^c}) \en
Note that the two configurations on
the RHS of (18) differ only at site $\alpha -\alpha_m$. We have then the
identity \be V_{\Lambda \alpha} ({ s}) = \sum_{m \geq 0}
\Phi^\Lambda_{Y^m_\alpha} ({ s}) + V_{\Lambda \alpha} ({ 0}) \en and we define
\be \Phi_{\Lambda X} ({ s}) = \Phi^\Lambda_{Y^m_\alpha} ({ s}) \en provided
there exists $m,\alpha$ such that $X=Y^m_\alpha$, otherwise we set
$\Phi_{\Lambda X} ({ s})=0$ (note that there is at most one such pair
$(n,\alpha)$). $\Phi_{\Lambda X}$ depends only on ${ s}_X$. Due to (16), (19),
we have \be {\cal H}_{\Lambda_N} ({ s}) = \sum_{X \subset \Lambda_N}
\Phi_{\Lambda X} ({ s}) + \mbox{constant} \en where constant = $\sum_{\alpha
\in \Lambda_N} V_{\Lambda \alpha} ({ 0})$. Note that (17) was chosen so that
$Y^m_\alpha$ with $|\alpha_m | < 1$ is a one-dimensional interval of length at
most $M$ in the time direction.

Now write \be \Phi_\Lambda = \Phi^0_\Lambda + \Phi^1_\Lambda \en where
$\Phi^0$ collects
the $X's$ with $| \alpha_m | < 1$ in (20) and $\Phi^1$ all the others. We
have the
following basic bound:

\begin{Pro} There exist $\epsilon_2 > 0$, $C< \infty$, such that, if $\epsilon
<
\epsilon_2$, then $\Phi_{Y^m_\alpha}$ given by (18) is bounded by \be |
\Phi^\Lambda_{Y^m_\alpha} ({ s}) | \leq C
\epsilon'^{\delta_{0i_m}}\lambda'^{t_m}
e^{-\beta' |i_m|} \en where $\alpha_m = (t_m,i_m)$ $\lambda' =
\lambda^{\delta/2}$,
$\epsilon' = \epsilon^\delta$, $\beta' =\frac{\beta \delta }{3}$ where
$\delta$ is as in
(3.4). Moreover, for any $\Lambda' \supset \Lambda$, \be |
\Phi^\Lambda_{Y^m_\alpha} ({
s}) - \Phi^{\Lambda'}_{Y^m_\alpha} ({ s})| \leq C \epsilon' e^{-\beta'
d(Y^m_\alpha ,
\partial \Lambda)} \en \end{Pro}

Choose now $M$ in (17) to be the smallest integer such that \be \lambda'^M <
\epsilon^\delta. \en Then, we have

\begin{Pro} There exist $\gamma > 0$, $\;\epsilon_0 > 0$, $C < \infty$ such
that, if
$\epsilon < \epsilon_0$, $\Phi^1$ satisfies the bound \be \sum_{\alpha \in
X} e^{\gamma
d(X)} \| \Phi^1_{\Lambda X} \| \leq C \epsilon^\delta \en uniformly in
$\alpha$ and
$\Lambda$. Moreover, \be | \Phi_{\Lambda X} ({ s}) - \Phi_{\Lambda' X} ({
s}) | \leq C
\epsilon^\delta e^{-\gamma d(X,\partial \Lambda)}. \en with $X \subset
\Lambda\subset
\Lambda'$. \end{Pro}

Since $\Phi^0$ has no coupling in spatial directions and has range $M$ in
the time
direction we get the following decomposition of its partition function
$Z^0$. We let for
$i\in {\bf Z}^d$ $J_i= ({\bf Z}_+\times i) \cap V$ so $V = \cup_i J_i$.
Then \be Z^0 (V |
{s}) = \prod_{i} Z^0 (J_i | s) \en

\begin{Pro} There exist $\gamma > 0$, $C < \infty$ independent of $M$ such
that, if $J $
is an interval of the form $[(i,t),(i,t+ \ell M)]$ then \be Z^0(J|s) =
\lambda^\ell W ({
s}_-) W(s_+) (1+ g_J (s_-,s_+)) \en where
$s_+=s_{((i,t+\ell M),(i,t+ (\ell +1)
M)]}$ and $s_-=s_{[(i,t- M),(i,t))}$ with \be | g_J (s_- , s_+) | \leq C
e^{-\frac{\gamma}{M}|J|} \en \end{Pro} Since this last Proposition is rather
standard in statistical mechanics, we defer its proof to the Appendix. For the
proof of the other Propositions, we need some Lemmas.

\begin{Lem} There exists $B < \infty$ such that $\psi_{\Lambda {\bf s}}$,
given by (11),
satisfies \BE | \psi_{\Lambda {\bf s}} ({\bf x})_i - \psi_{\Lambda' {\bf
s'}} ({\bf x}')_i
| \leq \lambda | x_i - x'_i| + B \epsilon (\sum_{j \in \Lambda}| x_j - x'_j
|  e^{-\beta
|i-j|} + e^{-\beta d(i,\partial \Lambda)}) \EN for $i \in \Lambda \subset
\Lambda'$, and
${ s}={ s}'|_\Lambda $ . For $\Lambda = \Lambda'$, we get (31) without the
term $
e^{-\beta d(i,\partial \Lambda)}$. \end{Lem}

\vspace*{5mm} \par\noindent {\bf Proof} Proceeding as in (5, 6), and using
Lemma 1 for
$a^{-1}_\Lambda$, we get $$ | a^{-1}_\Lambda ({\bf x})_i -
a^{-1}_{\Lambda'} ({\bf x}')_i
| \leq | x_i - x'_i | + B \epsilon (\sum_{j \in \Lambda}  | x_j - x'_j |
e^{-\beta |i-j|}
+ e^{-\beta d(i,\partial \Lambda)}) $$ Now, by (3.1), $(f^{-1})' < \lambda$
and this
implies the claim. \hfill$
\makebox[0mm]{\raisebox{0.5mm}[0mm][0mm]{\hspace*{5.6mm}$\sqcap$}}$ $ \sqcup$

\vspace*{5mm}

Iterating (31) then yields

\begin{Lem} Let $\psi_{\bf s} \equiv \psi_{\Lambda {\bf s}}$, $\psi'_{\bf
s'} \equiv
\psi_{\Lambda' {\bf s'}}$. Then, there exist $C <\infty$ such that \be &&|
\psi_{{\bf
s}_{t_n}} \cdots \psi_{{\bf s}_{t_1}} ({\bf x})_i - \psi'_{{\bf s'}_{t_n}}
\cdots
\psi'_{{\bf s'}_{t_1}} ({\bf x}')_i | \non\\ &&\leq \tilde \lambda^n (|x_i
- x'_i| + C
\epsilon \sum_{j \in \Lambda} e^{-\frac{\beta}{2}|i-j|} |x_j-x'_j|) + C
\epsilon
e^{-\frac{\beta}{2}d(i,\partial \Lambda)} \en where $i \in \Lambda \subset
\Lambda'$, ${
s}={ s}'|_\Lambda $ and $\tilde \lambda = \lambda^{1/2}$. For $\Lambda =
\Lambda'$, we get
(32) without the term $C \epsilon e^{-\frac{\beta}{2}d(i,\partial
\Lambda)}$. \end{Lem}

\vspace*{5mm} \par\noindent {\bf Proof of Lemma 4} The proof is done by
induction. For
$n=1$, we can use Lemma 3. Using (31) and (32) for $n-1$, we get \bea
&&|\psi_{{\bf
s}_{t_n}} \cdots \psi_{{\bf s}_{t_1}} ({\bf x})_i - \psi'_{{\bf s'}_{t_n}}
\cdots
\psi'_{{\bf s'}_{t_1}} ({\bf x}')_i | \\ &&\leq \lambda \; \tilde \lambda^{n-1}
(|x_i-x'_i| + C \epsilon \sum_{j \in \Lambda} e^{-\frac{\beta}{2}|i-j|}
|x_j-x'_j|) + C
\epsilon \lambda e^{-\frac{\beta}{2}d(i,\partial \Lambda)} \\ &&+ B
\epsilon \left[\sum_{j
\in \Lambda} \tilde\lambda^{n-1} |x_j -x'_j| e^{-\beta|i-j]} + C \epsilon
\tilde
\lambda^{n-1} \sum_{k,j \in \Lambda} e^{-\frac{\beta}{2}|j-k|} |x_k - x'_k |
e^{-\beta|i-j|} \right.\\ &&\left. + C\epsilon \sum_{j \in \Lambda}
e^{-\frac{\beta}{2}
d(j,\partial \Lambda)} e^{-\beta |i-j|}\right] + B \epsilon e^{-\beta
d(i,\partial
\Lambda)} \eea Regrouping terms, and exchanging $k,j$ in $ \sum_{k,j}$, we
get \bea &\leq&
\lambda \; \tilde \lambda^{n-1} |x_i - x'_i| \\ &+& (\lambda \; \tilde
\lambda^{n-1} C
\epsilon + B \epsilon \; \tilde \lambda^{n-1} + B C \epsilon^2 \; \tilde
\lambda^{n-1}
\sum_{k \in \Lambda} e^{-\frac{\beta}{2} |i-k|}) \sum_{j \in \Lambda}
e^{-\frac{\beta}{2}
|i-j|} |x_j-x'_j| \\ &+& (C \epsilon \lambda + B C \epsilon^2 \sum_{j \in
\Lambda}
e^{-\frac{\beta}{2}|i-j|} + B \epsilon) e^{-\frac{\beta}{2} d(i,\partial
\Lambda)} \eea
Now choose $C$ large enough so that $\lambda C + \frac{B}{\lambda} + {\cal
O}(\epsilon) <
\tilde \lambda C$ (which is possible since $\tilde \lambda = \lambda^{1/2}
> \lambda$) and
we get (32) for $n$. \hfill$
\makebox[0mm]{\raisebox{0.5mm}[0mm][0mm]{\hspace*{5.6mm}$\sqcap$}}$ $ \sqcup$

\vspace*{5mm} \par\noindent {\bf Remark} It is easy to extend Lemmas 3 and
4 to the
situation where $\Lambda = \Lambda'$ but $s \neq s'$ and to get,
instead of (32), \be
&|&\psi_{{\bf s}_{t_n}} \circ \cdots \circ \psi_{{\bf s}_{t_1}} ({\bf x})_i -
\psi_{{\bf s}'_{t_n}} \circ \cdots \circ \psi_{{\bf s}'_{t_1}} ({\bf
x}')_i| \nonumber\\
 &\leq& \tilde \lambda^n (|x_i - x'_i| + C \epsilon \sum_{j
\in \Lambda} e^{-\frac{\beta}{2} |i-j|} |x_j - x'_j |) \nonumber\\
&+&
\sum^n_{k=1} \tilde \lambda^{(n-k+1)} (| s_{(t_k,i)} - s'_{(t_k,i)} | + C
\epsilon \sum_{j \in \Lambda} e^{- \frac{\beta}{2} |i-j|} |s_{(t_k,j)} -
s'_{(t_k,j)} |) \en

\vspace*{5mm} \par\noindent {\bf Proof of Proposition 1} Using (15) and
(18) we can write
$$ \Phi^\Lambda_{Y^m_\alpha} ({ s}) = g(\psi_{{\bar{\bf s}}_t} \circ \cdots
\circ
\psi_{{\bar{\bf s}}_l}  ({\bf y})) - g (\psi_{{\bar{\bf s}}_t}\circ \cdots
\circ
\psi_{{\bar{\bf s}}_l} (\bar{\bf y})) $$ where $\alpha =(t,i)$,
$l=t-t_m+1$, ${\bar{\bf
s}}_{t_k}$ ,${t_k}=t,\dots, l$ is a configuration that coincides with ${\bf
s}_{t_k}$ in
$Y^m_\alpha$ and with $0$ outside, ${\bf y}=\psi_{{\bar{\bf s}}_{t-t_m}}
\circ \psi_{{\bf
0}} \circ \cdots \circ \psi_{{\bf 0}} ({\bf 0})$ and $\bar{\bf y}$ is
defined similarly,
but with ${\bar{\bf s}}_{t-t_m}$ coinciding with ${\bf s}_{t-t_m}$ in
$Y^{m-1}_\alpha$ and
with $0$ outside, so that $\bar{ y}_j = y_j$ for $|i-j| < |i_m|$. Finally,
$g$ is given
by (15).

Using the H\"older continuity of $f'$ and Lemma 2, we get \bea &&|
\Phi^\Lambda_{Y^m_\alpha} ({ s})| \leq C |(\psi_{{\bar{\bf s}}_t} \circ
\cdots \circ
\psi_{{\bar{\bf s}}_l} ({\bf y}))_i - (\psi_{{\bar{\bf s}}_t} \circ \cdots
\circ
\psi_{{\bar{\bf s}}_l}  (\bar{\bf y}))_i|^\delta \\ && + C \epsilon \sum_{k
\in \Lambda}
e^{-\frac{\beta}{2}|i-k|} |(\psi_{{\bar{\bf s}}_t} \circ \cdots \circ
\psi_{{\bar{\bf
s}}_l} ({\bf y}))_k - (\psi_{{\bar{\bf s}}_t} \circ \cdots \circ
\psi_{{\bar{\bf s}}_{l}}
(\bar{\bf y})_k |^\delta \eea

Now, use (32) in Lemma 4 for $\Lambda = \Lambda'$ and the fact that $\bar{
y}_j = y_j$ for
$|i-j| < |i_m |$ to get (23). By definition, ${\tilde \lambda}^\delta =
\lambda^{\delta/2}
=\lambda'$.

For (24), we bound \bea &&| \Phi^\Lambda_{Y^m_\alpha} ({ s}) -
\Phi^{\Lambda'}_{Y^m_\alpha} ({ s})| \\ &&\leq | g(\psi_{{\bar{\bf s}}_t}
\circ \cdots
\circ \psi_{\bar{\bf s}_l} ({\bf y})) - g(\psi'_{{\bar{\bf s}'}_t} \circ
\cdots \circ
\psi'_{{\bar{\bf s}'}_l} ({\bf y}))| \\ && + | g(\psi_{{\bar{\bf s}}_t}
\circ \cdots \circ
\psi_{\bar{\bf s}_l} (\bar{\bf y})) - g (\psi'_{{\bar{\bf s}'}_t} \circ
\cdots \circ
\psi'_{{\bar{\bf s}'}_l} (\bar{\bf y}))| \eea where $\psi'_{{\bar{\bf s}}'} =
\psi_{\Lambda' {\bar{\bf s}}'}$ and ${\bar{\bf s}'}_t$ is defined like
${\bar{\bf s}}_t$,
but with $\Lambda'$ instead of $\Lambda$. Now use the H\"older continuity
of $ f'$ and
Lemma 2 and then, use (32), where only the last term contributes. \hfill$
\makebox[0mm]{\raisebox{0.5mm}[0mm][0mm]{\hspace*{5.6mm}$\sqcap$}}$ $ \sqcup$

\vspace*{5mm} \par\noindent {\bf Proof of Proposition 2} The bounds (26),
(27) follow from
Proposition 1 because we excluded from $\Phi^1$ the terms with
$\delta_{0i_m} = 0$ and
$t_m < M$, and because of our choice of $M$ in (25): To get (26), we use
$d(X) \leq C(t_m
+ |i_m|)$ which follows from (20) and from the definition of $Y^m_\alpha$.
\hfill$
\makebox[0mm]{\raisebox{0.5mm}[0mm][0mm]{\hspace*{5.6mm}$\sqcap$}}$ $
\sqcup$ \vspace*{5mm}
\par\noindent
{\bf Proof of Theorem 3} We shall construct $\mu$ as the weak limit of
$T^{\ast N}_\Lambda  m_\Lambda = (P^N_\Lambda 1) m_\Lambda$, as $N \to
\infty, \Lambda
\uparrow {\bf Z}^d$, and, for that, we shall use the results of Section 2, the
representations (13, 21, 22) and the bounds of Proposition 2. Thus, let $G
\in L^\infty
({\cal M}_{B})$ , $|B|<\infty$ ; we have \be \int G \circ T^N_\Lambda
dm_\Lambda =
\sum_{({\bf s}_i)^N_1} \int_{C_\Lambda} G({\bf x}) \exp (-{\cal
H}_{\Lambda_N} ({ s}))
d{\bf x} \en We shall first construct the unique Gibbs state $\bar \mu$
with Hamiltonian
$\cal H$ on the following ``mixed" phase space: the lattice is
${\bf Z}^{d+1}_+$,
$\Omega_\alpha = S^1$ if $\alpha = (0,i)$, $i\in {\bf Z}^d$ and $\Omega_\alpha
=
\{0,\dots,k-1\}$ if $\alpha = (t,i)$, $t>0$. Then, $\mu$ will be the
restriction of $\bar
\mu$ to the ``time zero" phase space $\cal M$, as shown by (34). Since we
want to use
Theorem 2.2 (or, rather a trivial extension of it to the present setting),
let us check
now that our system satisfies the hypotheses of this theorem. We cover
${\bf Z}^{d+1}_+$
with $L$-cubes, $L = \ell_0 M$, where \be M e^{-\gamma \ell_0} \leq
\epsilon^\delta \en
(here, it would be more natural to take, instead of cubes, segments
parallel to the time
axis, but we shall keep the notations of Section 2). We bound $\|
\Phi^1_\Lambda \|_2 \leq
\epsilon$ uniformly in $\Lambda$ and we get (2.33) from Proposition 2 (with
$\epsilon = C
\epsilon^{\delta}$ of Proposition 2). We get (2.13,2.14) from (28) and
Proposition 3 as follows. Let $f = \log \lambda$ and $\phi_Y = \log (1+g_J)$
for $Y$ being and interval $J$ as in Proposition 3 with $|J|=nL$, $n\geq 1$
and $\phi_Y = 0$ otherwise. Then (30,35) imply (2.14) (with $\epsilon$
replaced by $C \epsilon^{\delta}$), using $$ \sum_{n \geq 1} Ln
e^{\frac{\gamma Ln }{L}} e^{- \frac{\gamma Ln }{M}} =M \sum_{n \geq 1} \ell_0
n e^{- \gamma (\ell_0 -1)n } \leq C M e^{-\gamma \ell_0}. $$

In order to be able to use estimates (2.30, 2.31) (see Remark 2.2), if we
observe that $L =
\ell_0 M$ with $M$ given by (25) and $\ell_0$ by (35) so that $M \leq C |
\log \epsilon |,
\ell_0 \leq C | \log \epsilon |$ and $L^{d+1} \epsilon^\delta << 1$ for
$\epsilon$ small.

So, Theorem 2.2 proves the existence of $\mu$. Its invariance under $T$ and
under the
lattice translations follows by construction: Let $G \in {\cal C} ({\cal
M}_B)$. Observe
that since the limit $\Lambda \uparrow {\bf Z}^d, N \to \infty$ taken above
can be taken
in any order, $$ \mu^\Lambda = \lim_{N \to \infty} (P^N_\Lambda 1)
m_\Lambda $$ exists and
is $T_\Lambda$ invariant. On the other hand, \be \lim_{\Lambda \uparrow
{\bf Z}^d} \| G
\circ T_\Lambda - G \circ T \|_{\infty} = 0 \en since $G$ is continuous.
Therefore, \be
\int G \circ T d \mu = \lim_{\Lambda \uparrow {\bf Z}^d} \int G \circ
T_\Lambda d \mu =
\lim_{\Lambda \uparrow {\bf Z}^d} \; \lim_{\Lambda' \uparrow {\bf Z}^d}
\int G \circ
T_\Lambda d\mu^{\Lambda'} \nonumber\\ = \lim_{\Lambda \uparrow {\bf Z}^d}
\lim_{\Lambda'
\uparrow {\bf Z}^d} \lim_{N \to \infty} \int G \circ T_\Lambda \circ
T^N_{\Lambda'}
dm_{\Lambda'} \en By (36), we may replace here $T_\Lambda$ by
$T_{\Lambda'}$ and we get $$
(37) = \lim_{\Lambda' \uparrow {\bf Z}^d} \lim_{N \to \infty} \int G \circ
T^{N+1}_{\Lambda'} dm_{\Lambda'} = \int G d\mu $$ The ${\bf
Z}^d$-translation invariance
of $\bar \mu$ implies the same invariance for $\mu$.

On the other hand, since the limit $\Lambda \uparrow {\bf Z}^d, N \to
\infty$ of (34)
exists for any $G \in L^\infty ({\cal M}_{B})$ with $|B|<\infty$, we see,
by taking
characteristic functions of sets of zero Lebesgue measure, that $\mu_\Lambda$
is
absolutely continuous with respect to the  Lebesgue measure, for any
$|\Lambda|<\infty$.

The space-time exponential mixing of $\mu$ follows essentially from the
exponential
clustering of $\bar{\mu}$ which itself follows from Theorem 2.2: By an
approximation
argument, it is enough, to prove (3.10), to consider $G\in {\cal C} ({\cal
M}_B)$, and $H$
as in (3.10): Then, as in (37),
\be &&\int G \circ T^n H d\mu - \int G d\mu \int H d \mu
\non\\ &&= \lim_{\Lambda \uparrow {\bf Z}^d} \; \lim_{N \to \infty}
(\int G P^n_\Lambda (H
P^N_\Lambda 1) dm_\Lambda - \int G P^N_\Lambda  1dm_\Lambda \int H P^N_\Lambda
1
dm_\Lambda). \en

Following equations (10,12,13), we get (replacing $H$ by
the periodic function
$h$):
 \be
P^n_\Lambda (h P^N 1) ({\bf x}) = \sum_{{\bf s}_1,\cdots,{\bf s}_{N+n}}
e^{-{\cal
H}_{\Lambda_{N+n}}({s})} h ({s}) \en where \be h({s}) \equiv h(\psi_{{\bf
s}_n} \circ
\cdots \circ \psi_{{\bf s}_1} ({\bf x})). \en

Proceeding as in (19), we write \be h({s}) = \sum_{m\geq 0} h_{Y^m_\alpha}
({s}) + \;
\mbox{constant} \en where $\alpha = (i,n)$ for some $i \in D$, and $$
h_{Y^m_\alpha} ({s})
= h ({s}_{Y^m_\alpha} \vee {0}_{(Y^m_\alpha)^c}) - h ({s}_{Y^{m-1}_\alpha}
\vee {0}_{(Y^{m-1}_\alpha)^c}) $$

Since $h$ is H\"older continuous, proceeding as in the proof of Proposition
1, we get an
estimate on $h_{Y^m_\alpha}$ similar to (23): \be \mid h_{Y^m_\alpha} (s) \mid
\leq C \| H \|_\delta  \lambda'^{tm} e^{-\beta'd(i-i_m, D)}
\en

Now, insert (39, 41) in (38), and use the exponential decay of correlation
(2.12) for $\bar \mu$ with $B$ being a subset of $\{0\} \times {\bf Z}^d$ and
$A = Y^m_\alpha$, so that
$d(A,B) \geq  d(i-i_m ,B) + (n-t_m)$, where the distance in the LHS is taken in
${\bf Z}^{d+1}_+$, while, on the RHS, it is taken in ${\bf Z}^{d}$.
Then, from (42) and
$$ \sum_{\ell \geq 0} \exp (-m(d(i-i_\ell, B)+ (n-t_\ell)))
\lambda'^{t_\ell} e^{-\beta'd(i-i_\ell, D)} \leq C e^{-m'(n+d(B,D))}
\min (|B|,|D|) $$ for
some $m' > 0$, we get (3.10).

Finally, let $\nu$ be a regular state. We want to show that $\forall B
\subset {\bf Z}^d,
|B| < \infty, \forall G \in {\cal C} ({\cal M}_B)$,
$$ \lim_{N \to \infty} \int G \circ
T^N d\nu = \int G d \mu . $$ Since
$\|G \circ T^N_\Lambda - G \circ T^N \|_\infty \to 0$, as
$\Lambda \uparrow {\bf Z}^d$, it is enough to show that \be \lim_{N \to \infty}
\lim_{\Lambda \uparrow {\bf Z}^d} (\int G \circ T^N_\Lambda d\nu - \int G
(P_\Lambda^N 1)
dm_\Lambda) = 0. \en Since $G \circ T^N_\Lambda$ is ${\cal B}_\Lambda$
measurable we may
replace $d\nu$ by $h_\Lambda ({\bf x}_\Lambda) d {\bf x}_\Lambda$ where
$h_\Lambda =
\frac{d\nu_\Lambda}{dm_\Lambda}$.

We have \be \int G \circ T^N_\Lambda h_\Lambda dm_\Lambda =
\int_{C_\Lambda} g P^N_\Lambda
h_\Lambda d{\bf x}_\Lambda \nonumber\\ = \sum_{({\bf s}_i)_{i=1}^N}
\int_{C_\Lambda} g
({\bf x}_B) \exp (- {\cal H}_{\Lambda_N} (s)) h_\Lambda (s) d{\bf
x}_\Lambda \en where
$h_\Lambda (s)$ is defined by (40), with $n$ replaced by $N$. With the same
notation, introduce $h_\Lambda (\bar s)$ where \be \bar s_\alpha = 0
\;\;\forall \alpha = (t,i),\;\;  t \leq \frac{N}{2} \;\; \mbox{and} \;\;
d(i,B) \leq N \en and $\bar s_\alpha = s_\alpha$ otherwise.
Let \be \bar
h_\Lambda (\bar s) = (\sum_{({\bf s}_i)^N_{i=1}} \int_{C_\Lambda} h_\Lambda
(\bar s) e^{-{\cal H}_{\Lambda_N} (s)} d {\bf x}_\Lambda)^{-1} h_\Lambda (\bar
s) \en Insert the identity
 \be h_\Lambda (s) = (1 -
\frac{\bar h_\Lambda (\bar s)}{h_\Lambda (s)}) h_\Lambda (s)
+ \bar h_\Lambda (\bar s) \en in (44).
Since \be \sum_{({\bf s}_i)^N_{i=1}} \int_{C_\Lambda} h_\Lambda (s) e^{{\cal
H}_{\Lambda_{N}}(s)} d{\bf x}_\Lambda = \int_{C_\Lambda} P^N_\Lambda h_\Lambda
d{\bf x}_\Lambda = \int_{C_\Lambda} h_\Lambda d{\bf x}_\Lambda = 1,
\en
(46) implies that
$\frac{\bar h_\Lambda (\bar s)}{h_\Lambda (s)}=
\frac{ h_\Lambda (\bar s)}{h_\Lambda (s)}
(\int \frac{ h_\Lambda (\bar s)}{h_\Lambda (s)}d \rho_\Lambda)^{-1}$,
where $d \rho_\Lambda$ is a probability measure. So,
we can bound \be | 1 - \frac{\bar h_\Lambda (\bar
s)}{h_\Lambda (s)} \mid \leq\exp (2 \sup_s | \log h_\Lambda (s) - \log
h_\Lambda (\bar s)|)-1. \en Now, since $\nu$ is regular
 \be | \log h_\Lambda
(s) - \log h_\Lambda (\bar s) | \leq C\sum_i | \psi_{{\bf s}_N} \circ \cdots
\circ \psi_{{\bf s}_1} ({\bf x})_i - \psi_{\bar{\bf s}_N} \circ \cdots \circ
\psi_{\bar{\bf s}_i} (\bar{\bf x})_i |^\delta \en Now we use (33) and the
definition (45) of $\bar s$ to bound each term in (50) by $C \lambda^{{\cal O}
(N)}$ if $d(i,B) \leq N$ and by $C \lambda^{{\cal O}(N)} e^{-\frac{\delta
\beta}{3} (d(i,B)-N)}$ if $d(i,B) > N$. Hence, (50) is bounded by $C (|B|)
\lambda^{{\cal O} (N)}$. Combining this with (49), (47), (48) and since $G$ is
bounded, we have \be &|& \int G \circ T^N_\Lambda d\nu - \int G (P^N_\Lambda 1)
dm_\Lambda | \nonumber \\ &\leq& |\sum_{({\bf s}_i)^N_{i=1}} \int_{C_\Lambda}
g ({\bf x}_B) \exp (- {\cal H}_{\Lambda_N} (s)) \bar h_\Lambda (\bar s) d {\bf
x}_\Lambda - \sum_{({\bf s}_i)^N_{i=1}} \int_{C_\Lambda} g ({\bf x}_B)
\exp (-{\cal H}_{\Lambda_N} (s)) d{\bf x}_\Lambda | \nonumber \\
&+& C(|B|)\| G
\|_\infty \lambda^{{\cal O}(N)} \en Since $g$ depends only on ${\bf x}_B$ and
$\bar h_\Lambda (\bar s)$ depends only on $s_\alpha$, with $\alpha = (t,i), t >
\frac{N}{2}$ or $d(i,B) > N$ the first term in (51) is also bounded by $C G
\lambda^{{\cal O}(N)}$, using the exponential decay of the Gibbs state defined
by ${\cal H}$ (uniformly in $\Lambda$ and $N$) proven above and the
normalization $$ \sum_{({\bf s}_i)^N_{i=1}} \int_{C_\Lambda} \exp (-{\cal
H}_{\Lambda_N} (s)) \bar h_\Lambda (\bar s) d{\bf x}_\Lambda = 1 $$ which
follows from (46). \hfill$
\makebox[0mm]{\raisebox{0.5mm}[0mm][0mm]{\hspace*{5.6mm}$\sqcap$}}$ $ \sqcup$

\vspace*{5mm} \par\noindent {\Large{\bf Appendix}}

\vspace*{5mm} \par\noindent {\bf Proof of Proposition 3} We need to study a
one-dimensional system with potentials $\Phi^0_Y$, $Y \subset {\bf Z}_+$, $Y =
[t_1,t_1+1,\cdots,t_2]$, $|t_2-t_1| < M$, $t_1 \geq 1$ satisfying $$ |
\Phi^0_Y | \leq C
\lambda'^{|Y|} \eqno{(A.1)} $$ for $\lambda'<1$ (we suppress here the $\Lambda$
dependence of $\Phi$).

Let ${\cal T}$ be the transfer matrix for $\Phi^0$, i.e. ${\cal T}$ is the
linear operator
on functions on $S \equiv \{0,1,\cdots,k-1\}^M$ (i.e. on ${\bf R}^{|S|}$)
given by $$
({\cal T}f)(s) = \sum_{t \in S} {\cal T} (s,t)f(t) \eqno{(A.2)} $$ with $$
{\cal T} (s,t)
= a(s)a(t)e^{U(s,t)} \eqno{(A.3)} $$ where $$ a(s) = \exp({\frac{1}{2}
\sum_{Y\subset
[1,M]} \Phi^0_Y(s)}) \eqno{(A.4)} $$ and $$ U(s,t) = \sum_Y \Phi^0_Y (s \vee t)
\eqno{(A.5)} $$ where $Y \subset [1,2M]$ with $M,M+1 \in Y$ and $(s \vee
t)_i$; equals
$s_i$ if $i\leq M$ and $t_{i-M}$ if $i>M$.

Let $J$ be as in the Proposition; we have $$ Z^0 (J|s) = (f_{s_-},{\cal
T}^\ell h_{s_+})
\eqno{(A.6)} $$ where $(\cdot,\cdot)$ is the scalar product in ${\bf
R}^{|S|}$, and $$
f_{s_-} (s) = a(s)e^{U(s_-,s)} $$ $$ h_{s_+} (t) = a(t) e^{U(t,s_+)}
\eqno{(A.7)} $$ We
can apply the Perron-Frobenius theory since ${\cal T}$ has strictly
positive entries.
${\cal T}$ has a non-degenerate largest eigenvalue $\lambda$ with left and
right
eigenvectors $u$ and $v$: $$ u {\cal T} = \lambda u $$ $$ {\cal T} v = \lambda
v
\eqno{(A.8)} $$ The vectors $v$ and $u$ can be chosen with positive
entries: $u(s), v(s)
>0$, $\forall s \in S$ and we normalize them by $$ \sum_{s\in S} u(s) v(s) = 1
>\eqno{(A.9)}
$$ We may write ${\cal T} = \lambda Q + R$ where $Q(s,t) = v(s) u(t)$, $Q^2
=Q$ and $QR =
RQ =0$.

Hence ${\cal T}^\ell = \lambda^\ell Q + R^\ell$, and $$ (f_{s_-} , {\cal
T}^\ell h_{s_+})
= \lambda^\ell (f_{s_-},v) (u,h_{s_+}) \Bigg(1+ \frac{(f_{s_-},R^\ell
h_{s_+})}{\lambda^\ell (f_{s_-},v)(u,h_{s_+})} \Bigg) \eqno{(A.10)} $$ So
define $W(s_-)
= (f_{s_-},v), W(s_+) = (u,h_{s_+}),$ and $$ g_J (s_- , s_+) =
\frac{(f_{s_-},R^\ell
h_{s_+})}{\lambda^\ell (f_{s_-},v)(u,h_{s_+})} \eqno{(A.11)} $$ so that,
using (A.6),
$Z(J|s_-,s_+)$ is of the form (4.29). To estimate $ g_J$ it is convenient
to introduce the
matrix $$ P(s,t) = \frac{{\cal T}(s,t)v(t)}{\lambda v(s)} \eqno{(A.12)} $$
Since $P(s,t) >
0, \sum_t P(s,t) = 1$, we can view $P(s,t)$ as the transition probability (from
state $s$
to state $t$) of a Markov chain. $P$ has a unique stationary distribution
$pP=p$, where
$p(s) = u(s) v(s)$. Write $$ P = \tilde Q + \tilde R \eqno{(A.13)} $$ where
$$ \tilde Q
(s,t) = \frac{Q(s,t) v(t)}{v(s)} = u(t)v(t) = p(t) \eqno{(A.14)} $$ and $$
\tilde R (s,t)
= \frac{R(s,t)v(t)}{\lambda v(s)} \eqno{(A.15)} $$ We have $\tilde Q \tilde
R = \tilde R
\tilde Q = 0$ and $P^\ell = \tilde Q + \tilde R^\ell$.

A standard result in the ergodic theory of Markov chains implies that $$
\sum_t | \sum_s
q(s) P^\ell (s,t) - p(t)| \leq 2(1-\gamma)^\ell \eqno{(A.16)} $$ where $$
\gamma =
\min_{s,s',t} \frac{P(s',t)}{P(s,t)}, \eqno{(A.17)} $$ for any $q$, with
$q(s) \geq 0$,
$\sum_s q(s) = 1.$  Postposing the proof of (A.16), we rewrite it, using
(A.13,A.14), as $$
\sum_t | \sum_s q(s) \tilde R^\ell (s,t) | \leq 2 (1-\gamma)^\ell
\eqno{(A.18)} $$ We use
(A.11,A.15) to write $$ | g_J (s_-,s_+)| = \Bigg| \frac{(f_{s_-}v,\tilde R^\ell
h_{s_+}v^{-1})}{(f_{s_-},v)(u,h_{s_+})} \Bigg| $$ $$ \leq 2(1-\gamma)^\ell
\frac{\max_t
h_{s_+} (t) v(t)^{-1}}{\min_s h_{s_+} (s) v(s)^{-1}}, \eqno{(A.19)} $$
using (A.18) with $$
q(s) = \frac{f_{s_-}(s) v(s)}{(f_{s_-},v)} = \frac{f_{s_-}(s)v(s)}{\sum_t
f_{s_-} (t)
v(t)}, $$ and using $\sum_s u(s)v(s) = 1$ to bound $(u,h_{s_+}) \geq \min_s
h_{s_+} (s)
v(s)^{-1}$.

Now, we need to show that $\gamma > 0$ uniformly in $M$ and we have to
bound the last
factor in (A.19) by a constant independent of $M$. Using (A.12,A.17) $$ \gamma
=
\min_{s,s',t} \frac{{\cal T}(s',t)v(s)}{{\cal T}(s,t)v(s')} = \min_{s,s',t}
\frac{{\cal
T}(s',t)\sum_r{\cal T}(s,r)v(r)}{{\cal T}(s,t)\sum_r {\cal T}(s',r)v(r)} $$
where we use
(A.8) in the last equality. So, since $v(r)>0$, $$ \gamma \geq
\min_{s',t,s,r} \frac{{\cal
T}(s',t){\cal T}(s,r)}{{\cal T}(s,t){\cal T}(s',r)} \geq e^{-4\|U\|} $$ by
(A.3). Using
(A.1,A.5), we see that $\|U\| \leq C_1 (\lambda')$, uniformly in $M$. Now
consider $$
\frac{h_{s_+}(t)v(s)}{h_{s_+}(s)v(t)} \leq \max_{s,t,r} \frac{h_{s_+}(t){\cal
T}(s,r)}{h_{s_+}(s) {\cal T}(t,r)} \leq e^{4\|U\|} \eqno{(A.20)} $$ by
(A.3,A.7).
Inserting (A.20) in (A.19) yields (4.30).

We are left with the proof of (A.16). Since $p P = p$ and since $\Sigma
p(s) = \Sigma q(s)
= 1$, the LHS of (A.16) can be written as $$\sum_t | \sum_{s,s'} (P^\ell
(s,t) - P^\ell
(s',t)) q(s) p(s') |, $$ and it is enough to prove that $$ \sum_t | P^\ell
(s,t) - P^\ell
(s',t)| \leq 2 (1-\gamma)^\ell \eqno{(A.21)} $$ for all $s,s'$. Let
$f_{s,s'} (t) = \;
\mbox{sign} \; (P^\ell (s,t) - P^\ell (s',t))$. Then, the LHS of (A.21)
equals \bea &&
|(P^\ell f) (s) - (P^\ell f) (s')| \\ &=& | \sum_z P(s,z) (P^{\ell-1}f)(z) -
P(s',z)(P^{\ell-1} f)(z)| \eea $$ \leq \frac{1}{2} (\sum_z |P(s,z) -
P(s',z)|) \sup_{z,z'}
|(P^{\ell-1} f) (z) - (P^{\ell-1} f) (z')| \eqno{(A.22)} $$ where, to get
the last
inequality, we use $$ |\mu_1 (F) - \mu_2 (F) | \leq \frac{1}{2} (\sum_z |
\mu_1 (z) -
\mu_2 (z) |) \sup_{z,z'}| F(z) - F(z')| $$ (see e.g. \cite{Si} Lemma 5.1.8)
where $\mu_1 ,
\mu_2$ are the probability measures $\mu_1 (z) = P(s,z)$, $\mu_2 (z) = P
(s',z)$ and $F =
P^{\ell-1} f$.

Since $P(s,z), P(s,z')$ are probability measures, $$ \frac{1}{2} \sum_z |
P(s,z) -
P(s',z)| = \sum_z{^+} (P (s,z) - P(s',z)) $$ where the last sum runs over
positive terms $$
= \sum_z{^+} (1- \frac{P(s',z)}{P(s,z)} ) P(s,z) \leq 1-\gamma
\eqno{(A.23)} $$ by
definition (A.17) of $\gamma$.

Inserting (A.23) in (A.22) and iterating yields (A.21), using in the last
step $$
\sup_{z,z'} | f_{s,s'} (z) - f_{s,s'} (z') | \leq 2. $$ \hfill$
\makebox[0mm]{\raisebox{0.5mm}[0mm][0mm]{\hspace*{5.6mm}$\sqcap$}}$ $ \sqcup$

\vspace*{3mm} \no{\Large\bf Acknowledgments}

\vs{3mm}

We would like to thank L. Bunimovich, A. van Enter, C. Maes and Y. Sinai
for interesting
discussions. This work was supported by NSF grant DMS-9205296, by EC grants
SC1-CT91-0695
and CHRX-CT93-0411 and was done in part during a visit of the authors to the
Mittag-Leffler Institute.

 \end{document}